\documentclass{nature}
\usepackage[T1]{fontenc}
\usepackage{amsmath,amssymb,graphics,epsfig,epstopdf,color,verbatim,ulem,braket,tabularx}
\usepackage{multirow}
\usepackage{booktabs}
\usepackage{makecell} 

\usepackage{listings}
\usepackage{cancel}
\usepackage{mathrsfs}
\usepackage{soul}
\usepackage{color}
\usepackage{url}
\usepackage{caption}

\usepackage{bm}
\usepackage{hyperref} 
\setstcolor{red}
\usepackage{lipsum}
\title{Correlated electronic structures and unconventional superconductivity in bilayer nickelate heterostructures}
\author{Changming Yue$^{1,2,3,*\#}$, Jian-Jian Miao$^{2,1,\#}$, Haoliang Huang$^{1,2,\#}$, Yichen Hua$^{1,\#}$, Peng Li$^{1,2}$, Yueying Li$^1$, Guangdi Zhou$^{1}$, Wei Lv$^{1}$, Qishuo Yang$^{1}$, Fan Yang$^{4}$, Hongyi Sun$^{2}$, Yu-Jie Sun$^{1,2}$, Junhao Lin$^{1,2}$, Qi-Kun Xue$^{1,2,5}$, Zhuoyu Chen$^{1,2,*}$, Wei-Qiang Chen$^{1,2,*}$}
\begin{document}

\maketitle

\begin{affiliations}
\item Department of Physics and Guangdong Basic Research Center of Excellence for Quantum Science, Southern University of Science and Technology, Shenzhen 518055, China
\item Quantum Science Center of Guangdong-Hong Kong-Macao Greater Bay Area, Shenzhen 518045, China
 \item Guangdong Provincial Key Laboratory of Advanced Thermoelectric Materials and Device Physics, Southern University of Science and Technology, Shenzhen, 518055, China
\item School of Physics, Beijing Institute of Technology, Beijing 100081, China
\item Department of Physics, Tsinghua University, Beijing 100084, China
\end{affiliations}
\leftline{$\#$These authors contribute equally.}
\leftline{$^*$Corresponding authors: yuecm@sustech.edu.cn, chenzhuoyu@sustech.edu.cn, chenwq@sustech.edu.cn}
\vspace*{1cm}

\begin{abstract} 
The recent discovery of ambient-pressure superconductivity in thin-film bilayer nickelates opens new possibilities for investigating electronic structures in this new class of high-transition temperature ($\bm{T_\mathrm{C}}$) superconductors. 
Here, we construct a realistic multi-orbital Hubbard model for the thin-film system, by integrating ab initio calculations with scanning transmission electron microscopy (STEM) measurements, which reveal a higher-symmetry lattice. The interaction parameters are calculated with the constrained random phase approximation (cRPA).
Density functional theory (DFT) plus cluster dynamical mean-field theory (CDMFT) calculations, with cRPA calculated on-site Coulomb repulsive $\bm{U}$ and experimentally measured electron filling $\bm{n}$, quantitatively reproduces Fermi surfaces from angle-resolved photoemission spectroscopy (ARPES) experiments.
The distinct Fermi surface topology from simple DFT+$U$ results features the indispensable role of correlation effects.
Based upon the correlated electronic structures, A modified random-phase-approximation (RPA) approach yields a pronounced $\bm{s^{\pm}}$-wave pairing instability, due to the strong spin fluctuations originated from Fermi surface nesting between bands with predominantly $\bm{d_{z^{2}}}$ characters.
Our findings highlight the quantitative effectiveness of the DFT+cRPA+CDMFT approach that precisely determines correlated electronic structure parameters without fine-tuning. The revealed intermediate correlation effect may explain the same order-of-magnitude onset $\bm{T_\mathrm{C}}$ observed both in pressured bulk and strained thin film bilayer nickelates.
\end{abstract}

\newpage
\section*{Introduction}
The high-temperature superconductivity under pressure in the Ruddlesden-Popper (RP) nickelates has garnered significant attention\cite{Wang327_Nature2023}. Extensive experimental\cite{hou2023emergence,Zhang_2024_HighTc,Xie_2024_MagneticExchange,Li_2024_ElectronicCorrelation,Wang_2024_La2PrNi2O7_Nature,dan_2024,wang2024pressure,cai_2024,liu_2024}
and theoretical\cite{YaoPRL2023,Yang_2023_sPlusMinus ,Liu_2023_sPlusMinus ,Yang_2023_ValenceBonds ,Qin_2023_HighTc ,zhang_2023,Christiansson_2023,pupha_l2023,liao_2024,zhang2024electronic,zhang2023trends,Tian_2024_Correlation ,Oh_2024_StrongPairing ,Fan_2024_BilayerCoupling ,RyeePRL2024CMDFT,wang2024electronic,chen2024electronic,lu2024interlayer,sakakibara2024,jiang2024high,zhang2024strong,Zheng_2025_TwoOrbital} studies have been conducted to investigate their physical properties and superconducting mechanisms. The high-pressure conditions required for the superconducting phase in bulk compounds pose significant challenges for detailed property measurements\cite{cai_2024}. Recently, the discovery of ambient-pressure superconductivity in bilayer nickelate epitaxial thin films grown on SrLaAlO$_4$ substrates open new possibilities for experimental investigations\cite{Ko_LNO327AmbientSC, Zhou_2024_AmbientSC,liu2025superconductivity}. Structurally, the out-of-plane (i.e. $c$-axis) lattice constants in compressively strained films are elongated, opposite to anisotropically pressured bulks despite their similar onset $T_\mathrm{C}$, raising an intriguing question about the role of the interlayer coupling via the $d_{z^{2}}$ orbital in superconductivity.

On the theoretical front, understanding the electronic structure is essential for exploring the superconducting mechanism of these materials. There has been considerable debate on the electronic structure and the role played by the correlation effect in superconducting bilayer nickelate bulks under high pressure\cite{zhang_2023,zhang2023trends,XJZhou_ARPES_327,Yang_2023_sPlusMinus,Li_2024_ElectronicCorrelation,wang2024electronic,jiang2024high}. The superconducting bilayer nickelate thin films under ambient pressure enables angle-resolved photoemission spectroscopy (ARPES) experiments that serve as a benchmark to address this issue\cite{LiPeng_ARPES_327_Film}. In this work, we study the correlated electronic structure and the pairing mechanism of bilayer nickelate superconducting thin films, based on structural parameters from experiments, and achieve quantitative agreements with ARPES Fermi surfaces without parameter fine-tuning.

\section*{Experimental Structural Analysis}
\begin{figure*}[htp]
\centerline{\epsfig{figure=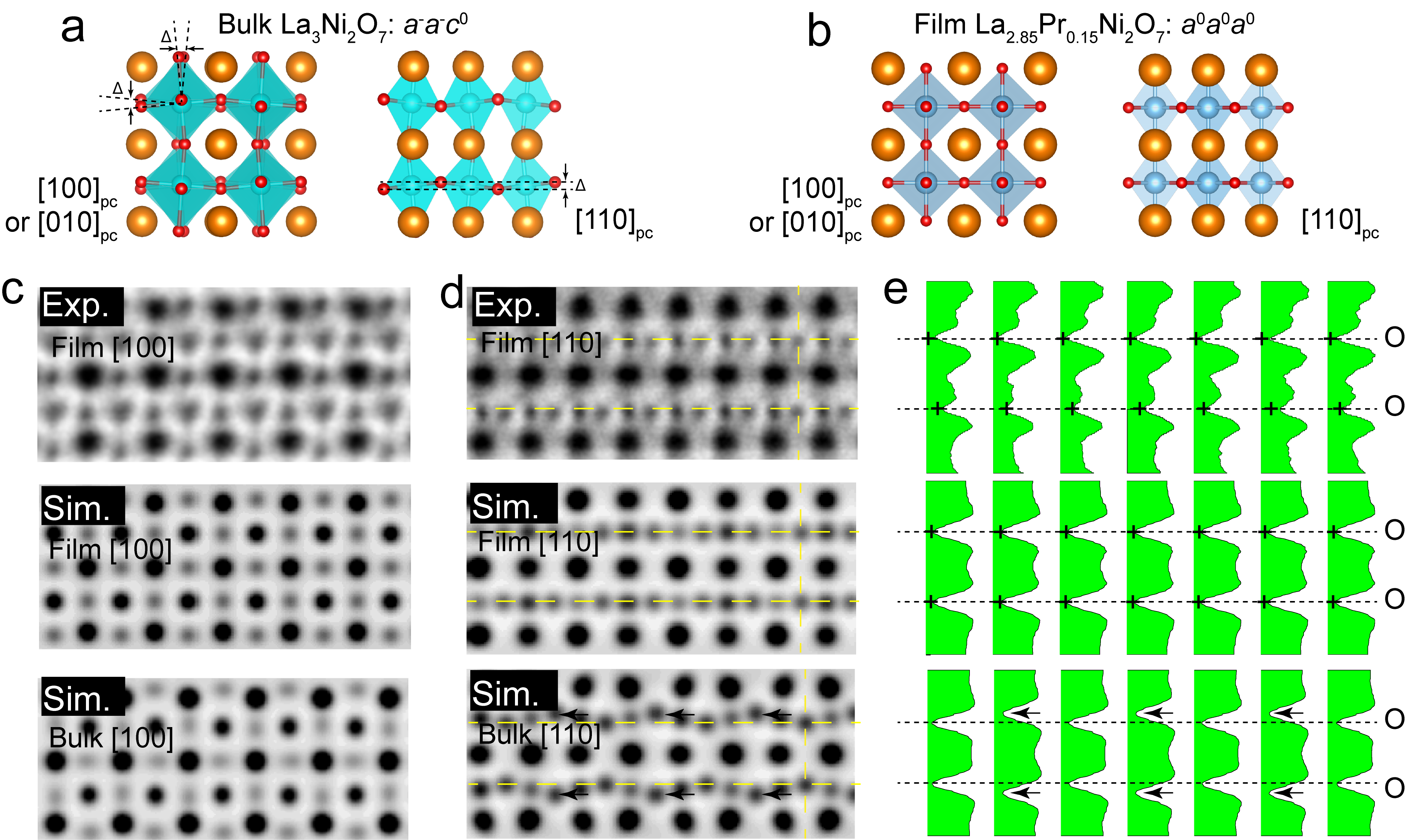,width=1.0\columnwidth}}
\caption{| {\bf Oxygen octahedral rotation analysis in La$_{2.85}$Pr$_{0.15}$Ni$_2$O$_7$ film}. {\bf a}, The schematic crystal structure of the bulk La$_3$Ni$_2$O$_7$ with an $a^-a^-c^0$ oxygen octahedral rotation pattern. Displacements of oxygen atoms are indicated by $\Delta$. {\bf b}, The schematic crystal structure of La$_{2.85}$Pr$_{0.15}$Ni$_2$O$_7$ film grown on SrLaAlO$_4$ substrate with an $a^0a^0a^0$ oxygen octahedral rotation pattern, showing no splitting or displacement in any direction. {\bf c} and {\bf d}, Zoom-in ABF images of the cross-section of La$_{2.85}$Pr$_{0.15}$Ni$_2$O$_7$ film, projected along [100]$_{pc}$, and [110]$_{pc}$, respectively. ABF images ([100]$_{pc}$, [010]$_{pc}$, and [110]$_{pc}$) with larger fields of view are shown in Fig. S1. From top to bottom are the experimental results of the film, the simulation results of the film, and the simulation results of the bulk. {\bf e}, Atomic columns intensity profiles in the project along [110]$_{pc}$ as the vertical dash line in panel {\bf d}. Crosses indicate the positions of the local minimum corresponding to the positions of oxygen atoms. Dashed lines mark the same positions as the horizontal dashed lines in panel {\bf d}. Black arrows indicate the positions of the oxygen atoms deviating from the dashed lines.
}
\label{fig:Oxygen_rotaion_detail}
\end{figure*}

Scanning transmission electron microscopy (STEM) analysis of oxygen octahedral rotations confirms a higher-symmetry lattice in the superconducting bilayer nickelate epitaxial films grown on SrLaALO$_4$ substrates\cite{Zhou_2024_AmbientSC}, than that in the ambient-pressure bulks. In La$_3$Ni$_2$O$_7$ bulk crystals without pressures, the $a^-a^-c^0$ oxygen octahedral rotation pattern is generally seen (Fig.~\ref{fig:Oxygen_rotaion_detail}{\bf a}): the NiO$_6$ octahedra undergo inverse rotations along the $a$ and $b$ axes (i.e. [100]$_{pc}$ and [010]$_{pc}$), while no rotation occurs along the $c$ axis ([001]$_{pc}$). Note that [100]$_{pc}$ and [010]$_{pc}$ are symmetric directions under the symmetry group. This rotation leads to splittings of oxygen positions along the [100]$_{pc}$/[010]$_{pc}$, and [110]$_{pc}$ projections. 
In contrast, the La$_{2.85}$Pr$_{0.15}$Ni$_2$O$_7$ film grown on SrLaAlO$_4$ appears more likely to conform to the high-symmetry $a^0a^0a^0$ pattern (Fig.~\ref{fig:Oxygen_rotaion_detail}{\bf b}). This pattern features the absence of splittings of the oxygen positions in any projection directions. 
Fig.~\ref{fig:Oxygen_rotaion_detail}{\bf c} and {\bf d} present zoomed-in annular bright-field (ABF) images of the La$_{2.85}$Pr$_{0.15}$Ni$_2$O$_7$ film cross-section, projected along the [100]$_{pc}$ and [110]$_{pc}$ directions, respectively (larger field of view images in Fig. S1). In the [100]$_{pc}$ direction, the oxygen atoms in both the experimental and simulated film results appear rounder compared to the elliptical shape observed in the simulated bulk. In the [110]$_{pc}$ direction, the oxygen atoms exhibit a distinct zigzag feature in the Ni-O layer for the bulk simulated case, which is not present in the film measurement and the film simulation. The line profiles of oxygen atoms (Fig.~\ref{fig:Oxygen_rotaion_detail}{\bf e}) more clearly demonstrate their alignment in contrast to the bulk simulations. These results suggest that the coherent epitaxy may suppress the oxygen octahedral rotations in the film, giving rise to a higher symmetry of the lattice, thus further altering orbital overlap and bonding interactions between Ni and O atoms. Detailed positions of each cation in the lattice are measured combining STEM and X-ray diffraction (XRD) experiments, as shown in (Table~\ref{tab:doped_film_paras}).

\section*{Ab initio calculations}
\begin{figure*}[htp]
\centerline{\epsfig{figure=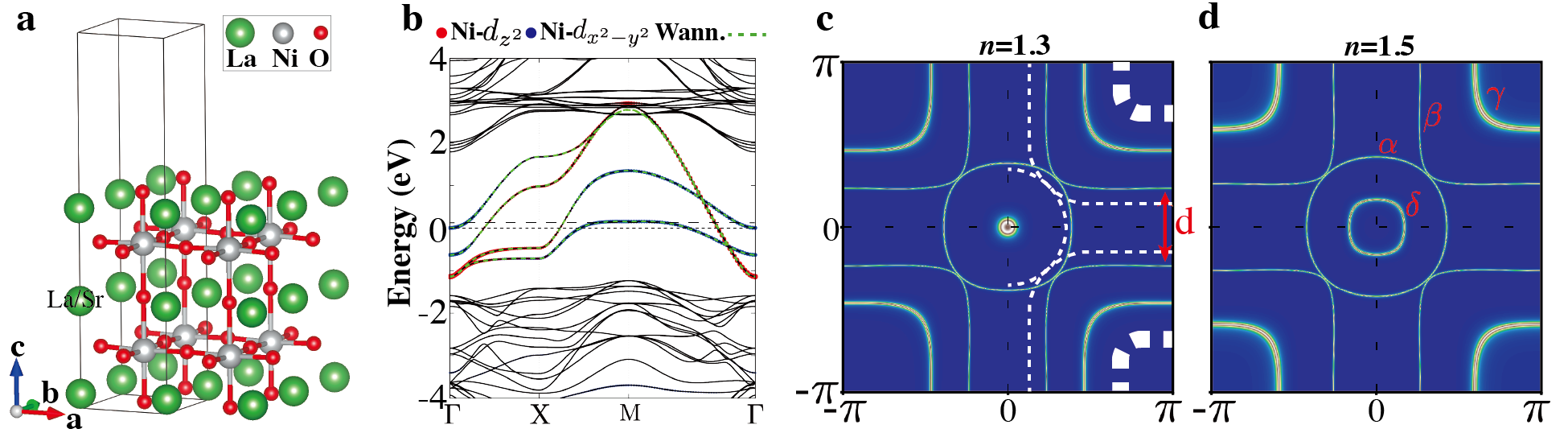,width=1.0\columnwidth}}
\caption{| {\bf Crystal Structure and DFT+U results (U=5, J=1).} 
{\bf a} The half-UC thin film crystal structure of La$_3$Ni$_2$O$_7$ constructed using structural parameters from the thin film of La$_{2.85}$Pr$_{0.15}$Ni$_2$O$_7$. 
{\bf b} The DFT+U bands (black solid line) and their Wannier interpolation (green dashed line) of the half-UC thin film. The thin (thick) dashed line marks the Fermi energy for the system with (without) Sr doping. The size of red (blue) dots demonstrates the projected weight for Ni-$d_{z^2}$ and Ni-$d_{x^2-y^2}$ orbitals. 
{\bf c-d} Fermi surfaces of Wannier TB model at $n$=1.3 and $n=$1.5 per Ni site, respectively. Names of all electron or hole pockets are labeled in {\bf c}.
In panel {\bf c}, the extracted ARPES Fermi surfaces\cite{LiPeng_ARPES_327_Film} is overlaid as the white dashed line. Only the right half is shown for clarity.
}
\label{fig:DFT_FS}
\end{figure*}

The crystal structure of thin films is constructed for DFT+$U$ calculations, using realistic structural parameters (Table~\ref{tab:doped_film_paras}). A simple yet effective choice is to use the half unit cell (UC) containing one nickelate bilayer (see Fig.~\ref{fig:DFT_FS}{\bf a}). 
Experiments show that Sr substitution of La near the interface due to interfacial diffusion from the substrate\cite{Zhou_2024_AmbientSC} may introduce moderate hole doping into the system. Such a Sr-doping effect is simulated by the virtual crystal approximation, and the resultant band structure can be approximated by the rigid-band shift of the Fermi level in the one without Sr. More details of DFT+$U$ calculations can be found in supplementary materials (SM.2). The band structure is shown in Fig.~\ref{fig:DFT_FS}{\bf b}. Similar to the bulk structure\cite{YaoPRL2023,XJZhou_ARPES_327}, the low-energy bands are mainly formed by the Ni-$e_g$ orbitals as demonstrated by the orbital-projected bands.

A realistic multi-orbital Hubbard model for Ni-$e_g$ subspace is formulated (see Method). The first part is the tight-binding model which has bands aligned well with the DFT+$U$ band (see Fig.~\ref{fig:DFT_FS}{\bf b}). 
The TB model Hamiltonian in ${\bf k}$-space has the same matrix structure as in Ref.~\cite{YaoPRL2023}
The onsite energy level $\epsilon_{x/z}$ and the hopping parameters are summarized in Table \ref{tab:tb_paras}. Here, we introduce the inter-layer bonding (+) and anti-bonding (-) orbital basis $d_{z_{\pm}}=(d_{Az}\pm d_{Bz})/\sqrt{2}$ and $d_{x_{\pm}}=(d_{Ax}\pm d_{Bx})/\sqrt{2}$ ($A,B$ the layer index, $x\equiv d_{x^2-y^2}$, $z\equiv d_{z^2}$).
Compared to the high-pressure bulk crystal, the thin film has two main differences in the TB model parameters that significantly affect the bands and Fermi surface topology. First, the inter-layer intra-$z$-orbital hopping $t_{\perp,\text{0.5UC}}^z$=-0.439 eV is nearly 30\% (0.2 eV) smaller in magnitude than $t_{\perp,\text{HP}}^z$=-0.635 eV. Second, 
the crystal-field splitting $\Delta E_{\text{0.5\text{UC}}}=\epsilon_{x}-\epsilon_{z}$ is enhanced to 0.519 eV, roughly $\sim$40 \% larger than  $\Delta E_{\text{HP}}=0.367$ eV. Smaller $t_{\perp}^z$ means smaller bonding-anti-bonding splitting, which leads to the coexistence of $\delta$ electron pocket (located at $\Gamma$ point) formed by the $z_-$ band and the $\gamma$-hole pocket (located at $M$ point) formed by $z_+$ band, as shown in Fig.~\ref{fig:DFT_FS}{\bf c-d} for both hole-doped (with 0.2 hole doping per Ni) and un-doped scenarios. This coexistence, however, does not happen in both high-pressure\cite{YaoPRL2023} and ambient pressure bulk crystal\cite{XJZhou_ARPES_327}. 

We now compare the Fermi surfaces obtained from DFT+$U$ and ARPES experiments at both undoped ($n=1.5$) and hole-doped ($n=1.3$, as estimated in ARPES experiments\cite{LiPeng_ARPES_327_Film}) scenario. Significant discrepancies between DFT+$U$ results and ARPES data are apparent. For instance, the distance $d$ (marked by the red arrow in Fig.~\ref{fig:DFT_FS}{\bf c}) between two Fermi surface branches of the $\beta$ band near $(\pi,0)$ point is much smaller in AREPS than that in DFT results. The $\gamma$ pocket around $(\pi,\pi)$ seen in ARPES is much smaller than that in the DFT+$U$, and its diffuse nature may indicate that the $d_{z^2}$ states are much less coherent than the $d_{x^2-y^2}$ states. We will demonstrate that these discrepancies arise from the incomplete treatment of electron correlations in the DFT+U approach.

The second part is the rotation-invariant Kanamori-type multi-orbital on-site interaction. Realistic intra-orbital Coulomb interaction $U_\mathrm{RPA}\approx$ 3.77 eV and Hund's coupling $J_{\mathrm{RPA}}\approx$0.56 eV (see Table~\ref{tab:cRPA}) are obtained according to the constrained random phase approximation (cRPA)\cite{cRPA2004},  which counts in the screening effect from the rest energy bands out of the low-energy $e_g$ subspace. 
As non-local interaction $V$ can screen $U$ further\cite{Christiansson_2023} while $J$ unscreened, we tune $U$ close to $U_\mathrm{RPA}$ with fixed $J=J_{\mathrm{cRPA}}$= 0.56 eV to mimic this screening effect and also systematically investigate the effect of $U$ on the correlated electronic structure. By comparing the electronic structure calculated from CDMFT and that of the ARPES experiment, we find that the best fitting happens at $U \sim 3.6$ eV (will be shown below). In this regime, CDMFT shows clear quasi-particle (QP) bands but with some small QP weight ($Z_{d_z^2} \sim 0.2 \ll 1$), which suggests that the system is away from a doped Mott insulator but still have a significant correlation effect. In other words, our CDMFT results indicate that the thin film is an intermediate correlated system.

\section*{CDMFT Studies}
The main results of CDMFT are summarized in Fig.\ref{fig:CDMFT}, which in detail demonstrate how the physical quantities like quasi-particle weight, occupancy, effective energy level, spectral function, and Fermi surfaces vary as one increases $U$. The calculations are performed at $T$=200 K and $n$=1.3 per site, which is close to the filling 1.28 estimated in the ARPES experiment\cite{LiPeng_ARPES_327_Film}. More results on doping and temperature dependence can be found in SM.3, especially the results at lower temperatures like $T=$ 50 K, which only have quantitative differences from the results at $T=$ 200 K. 

As one increases $U$, we observe an increase in effective energy splitting
$\epsilon^{\text{eff}}_{z_-}-\epsilon^{\text{eff}}_{z_+}$ between the $z_+$ and $z_-$ orbitals as $U$ increases, where $\epsilon^{\text{eff}}_{\alpha} = \epsilon_{\alpha}+\mathrm{Re}\Sigma{\alpha(i0^+)} - \mu$. 
Thus the effective energy level $\epsilon^{\text{eff}}_{z_-}$ ($\epsilon^{\text{eff}}_{z_+}$) of the $z_-$ ($z_+$ ) orbital is pushed upward (downward) as seen in Fig.~\ref{fig:CDMFT}{\bf a}.
For $U\le$ 3.0 eV, $\epsilon^{\text{eff}}_{z_\pm}$ deviates slowly from its bare value $\epsilon_{z_\pm}$ (dashed lines). 
The deviation increases rapidly when $U\gtrsim 3.6$ eV, as indicated by the arrows. This is consistent with the picture of strong inter-layer electronic correlations between the $z$ orbitals\cite{RyeePRL2024CMDFT}, which becomes more prominent as $U$ increases. On the contrary, $\epsilon^{\text{eff}}$ for the $x$ orbitals remains gradually increasing even for larger $U$. $U=3.6$ eV could be regarded as a characteristic interaction strength where the correlation effect becomes significant. 

The $U$-dependent orbital occupancy per spin is given in Fig.~\ref{fig:CDMFT}{\bf c}. There is a self-doping effect with the electrons transfer from the $x$ orbitals to the $z$ orbitals (note that $n_d$=1.3 fixed). As $U$ is tuned from 2 to 3.77 eV, the total filling in the $z$ orbitals increases from $\sim$0.76 to $\sim$0.9 while the  total filling in the $x$ orbitals decreases from $\sim$0.54 to $\sim$ 0.4. The quasi-particle weight $Z$ of the $x_{\pm}$ orbitals ranges from $\sim$0.7 to $\sim$0.5 as $U$ is tuned down, which results from a small filling per spin ($\sim$ 0.3 to $\sim$0.2). The $Z$ of $z_\pm$ orbitals decreases from $\sim$0.45 to $\sim 0.2$ for the $z_+$ orbital and to $\sim$ 0.08 for $z_-$ orbital, showing that $z$ orbitals are much more correlated than the $x$ orbitals. The reason for this is that the self-doping effect from $x$ to $z$ orbital makes the filling in each $z$ orbital approach to half filling per spin (from $\sim$0.38 at $U$=2 to $\sim$0.45 at $U=3.77$.)

Associated with the enhanced effective level splitting as $U$ increases, the ${\bf k}$ spectral functions $A({\bf k},\omega)$ and Fermi surfaces evolve correspondingly. From panel \ref{fig:CDMFT}{\bf e-h}, one can see that both $\delta$-band at $\Gamma$ point and $\gamma$-band at $M$-point at low-energy become more flat and gradually fade away (being less coherent) from the Fermi energy. First, as mentioned before, there is a downward or upward shift of $\epsilon^{\text{eff}}_{z_\pm}$. Second, the quasi-particle spectral weight (see Fig.\ref{fig:CDMFT}{\bf b}) decreases, which makes especially the $\gamma$ band more flat near the Fermi energy. Furthermore,   the scattering rate is enhanced when $U$ increases (not shown). As a result, the Fermi surface of $\delta$ electron pocket gradually disappears, as seen from Fig.\ref{fig:CDMFT}{\bf i-l }. Simultaneously, the $\gamma$ hole pocket also shrinks in size and becomes less coherent (as evidenced by its decreased intensity in colormap). At $U=3.77$, the $\gamma$ band becomes flat near $M$ point, and its band-top touches the Fermi energy, which gives rise to the Fermi surface shown in Fig.~\ref{fig:CDMFT}{\bf n}. 
The $A({\bf k},\omega)$ and Fermi surfaces at $U=3.6$ to $U=3.77$ eV shown in Fig.~\ref{fig:CDMFT} quantitatively match all the key features seen in the ARPES Fermi surfaces, as indicated by the white dashed lines in Fig.~\ref{fig:CDMFT}{\bf l}. (Doping and temperature dependence of the physical quantities in CDMFT are presented in SM3.). The agreement with ARPES data originates from the enhanced effective level splitting for $U\gtrsim3.6$ eV, suggesting that the thin film is in an intermediate coupling regime.
 


\section*{Pairing Mechanism}
The superconducting pairing symmetry is studied with a modified RPA approach, where the quasiparticle Hamiltonian from the CDMFT is taken as an input (see {\bf Method} and SM.4 for details).  We assume that effective interaction used in RPA is in the same form as the one in Eq.~\ref{eq:Hloc} and we fix $\ensuremath{J_{\textrm{eff}}} = \ensuremath{U_{\textrm{eff}}}/6$ in the calculations.
The effective interaction strength $\ensuremath{U_{\textrm{eff}}}$ is a tuning parameter. It could be regarded as the strength of some kinds of residual interactions renormalized by a factor $Z^2$, where $Z$ is the quasi-particle weight.  Since $Z$ of $d_{z^2}$ orbital is quite small ($\sim 0.18$ for $U = 3.6$ eV), the effective interaction strength $\ensuremath{U_{\textrm{eff}}}$ should also be very small.

The RPA results for $U = 3.6$ eV and $n = 1.3$ are summarized in Fig.~\ref{fig:fig_RPA}.  In Fig.~\ref{fig:fig_RPA}{\bf a}, we show the typical RPA-renormalized spin susceptibility. The dominant wavevector $\mathbf{Q}_{1}$ corresponds to the nesting between the $\beta$-pocket and $\gamma$-pocket as shown in Fig.~\ref{fig:fig_RPA}{\bf c}.  In Fig.~\ref{fig:fig_RPA}{\bf b}, we show the dependence of effective pairing strength\cite{Scalapino_1986} $\lambda$ on $\ensuremath{U_{\textrm{eff}}}$ for various pairing symmetries.  It is clear that larger $\ensuremath{U_{\textrm{eff}}}$ leads to stronger superconducting instability. The leading pairing symmetry is always a $s^\pm$-wave, where the gap function on the Fermi surface is depicted in Fig.~\ref{fig:fig_RPA}{\bf c}.  
The coincidence of $\mathbf{Q}_{1}$ in Fig.~\ref{fig:fig_RPA}{\bf a} and the one connects the largest gap in magnitude in {\bf c} supports the spin-fluctuations-mediated pairing mechanism.
In real space, the strongest pairing is between the inter-layer $d_{z^{2}}$ orbitals as shown in Fig.~\ref{fig:fig_RPA}{\bf d}, which is consistent with the fact that the Fermi surface patches connected by $\mathbf{Q}_{1}$ on $\beta$ and $\gamma$ bands shown in Fig.~\ref{fig:fig_RPA}{\bf c} are mainly formed by $d_{z^{2}}$ orbitals.
We also perform similar calculations for $3\leq U\leq3.77$, and $s^{\pm}$-wave pairing is always dominant (see SM.4 for more details).  All in all, the most possible pairing symmetry is $s^{\pm}$-wave pairing for realistic parameters.


\section*{Discussion}
First, we discuss the extent of correlation effects in the bilayer nickelate films. The on-site Coulomb repulsion from cRPA is $U \sim 3.77$ eV, consistent with the typical value of $U = 3-4$ eV for the transition metals. Due to the strong interlayer coupling, the multi-orbital tight-binding model involves four bands with a total bandwidth around 4 eV, which is comparable to the on-site Hubbard $U$. The situation is similar to the one in the iron-based superconductor\cite{kamihara2008iron,qazilbash2009electronic,zhou2010electron,yi2013observation}, suggesting that the bilayer nickelate systems also lie in an intermediate correlation regime.

\begin{figure*}[htp]
\centerline{\epsfig{figure=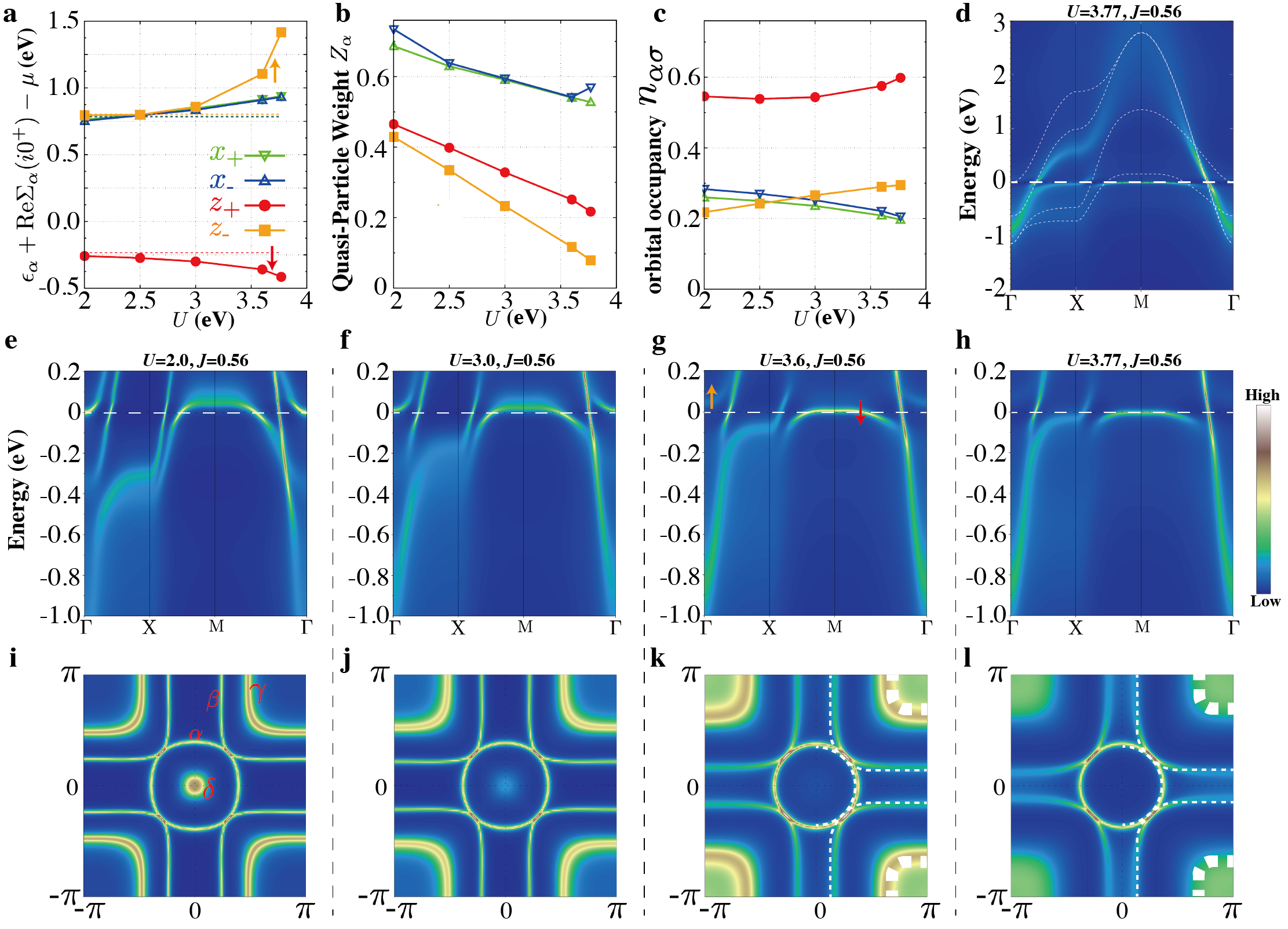,width=1.0\columnwidth}}
\caption{| {\bf $U$ dependence of CDMFT results at temperature $T$=200 K, filling $n=1.3$ per Ni (in $e_g$ orbitals) and Hund coupling $J=0.56$ eV.} 
{\bf a} The effective energy level $\epsilon^{\text{eff}}_{\alpha} = \epsilon_{\alpha}+\mathrm{Re}\Sigma{\alpha(i0^+)} - \mu$ for each bonding and anti-bonding orbital as a function of $U$ ($\mu$ is the chemical potential). The dashed lines mark the bare energy level $\epsilon_{\alpha}$ from DFT. 
{\bf b} The quasi-particle spectral weight $Z_{\alpha}$ (degenerate in spin). 
{\bf c} The orbital occupancy per spin $n_{\alpha\sigma}$.
{\bf d-h} The momentum-resolved spectral function $A({\bf k},\omega)$ at indicated $U$. We show  $A({\bf k},\omega)$ in large energy window $-2<\omega<3$ for $U=3.77$ in {\bf d} and in zoomed-in energy window $-1<\omega<0.2$ for $U=2.0$, $3.0$, $3.6$ and $3.77$ in ({\bf e-h}), respectively.
The white thin dashed line in {\bf d} indicates the DFT bands of Ni-$e_g$ orbitals. The thick dashed lines in {\bf d-h} mark the Fermi energy. 
{\bf i-l} The corresponding Fermi surface $A({\bf k},0)$ at indicated $U$  is shown in {\bf e-h}. 
In panel {\bf k-l}, the extracted ARPES Fermi surfaces\cite{LiPeng_ARPES_327_Film} is overlaid as the white dashed line. Only the right half is shown for clarity.
}
\label{fig:CDMFT}
\end{figure*}

\begin{figure*}
\centerline{\epsfig{figure=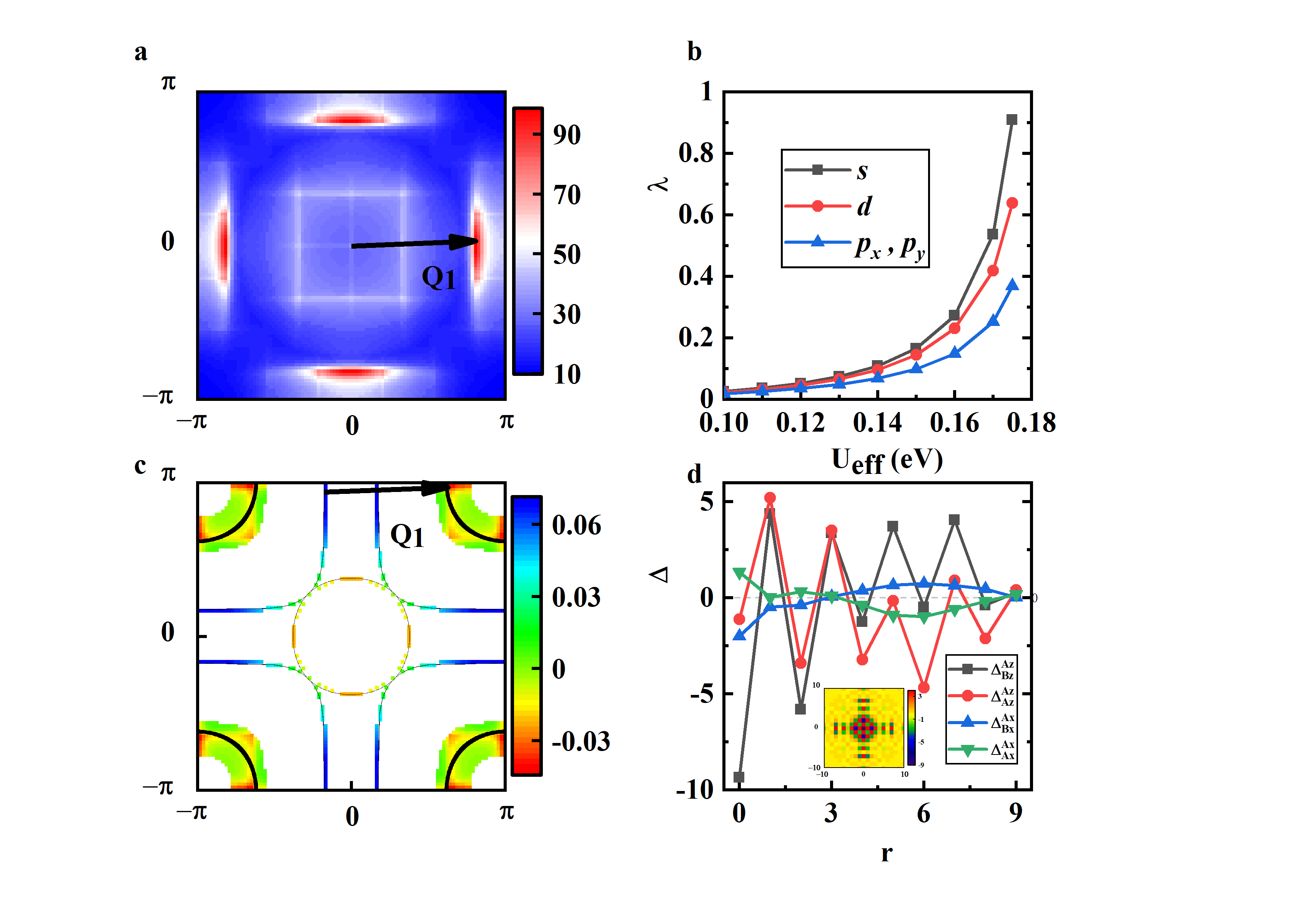,width=1.0\columnwidth}}
\caption{| {\bf The CDMFT+RPA calculation results on $\bm{96\times96}$ square lattice at fixed temperature $\bm{T=0.001}$ eV, filling $\bm{n=1.3}$ per Ni-$\bm{e_g}$ and bare interaction strength $\bm{U=3.6}$ eV.} 
{\bf a} The distribution of the largest eigenvalue $\chi^{\left(s\right)}\left(\mathbf{q}\right)$ of the RPA-renormalized spin susceptibility matrix in the Brillouin zone for $\ensuremath{U_{\textrm{eff}}}=0.17$ eV. $\mathbf{Q}_{1}$ is the representative wave vector of eight equivalent distribution peaks related by $D_{4h}$ point group. {\bf b} The largest eigenvalue $\lambda$ of the effective pairing interaction vertex matrix as a function of the effective interaction strength $U_{\textrm{eff}}$ for leading $s$, $d$ and degenerate $p_{x}$, $p_{y}$-wave pairings. {\bf c} The distribution of the leading $s^{\pm}$-wave pairing gap function near Fermi surface for $\ensuremath{U_{\textrm{eff}}}=0.17$ eV. The black thin lines denote the Fermi surface. The wave vector $\mathbf{Q}_{1}$ in {\bf a} and Fermi surface nesting vector in {\bf c} are identical. {\bf d} The real space $\mathbf{r}$ dependence of the $s^{\pm}$-wave pairing gap function in the orbital basis for $\ensuremath{U_{\textrm{eff}}}=0.17$ eV. $\mathbf{r}$ is the distance in unit cell between the two orbitals involved in the pairing. The inset shows the spatial distribution of $\Delta_{Bz}^{Az}$. $\Delta_{\tau'\gamma'}^{\tau\gamma}$ denotes the pairing between $\gamma$-orbital on $\tau$-layer and $\gamma'$-orbital on $\tau'$-layer. The other input parameters are all from CDMFT.
}
\label{fig:fig_RPA}
\end{figure*}

Although the correlation is moderate, its effects are non-negligible. The weak-coupling fluctuation-exchange (FLEX) approximation\cite{bickers1989,witt2021,sakakibara2024} at the same filling $n=1.3$ fails to reproduce ARPES Fermi surfaces.
As a weak coupling approach, FLEX calculations are restricted to $U\le$1.8 eV, beyond which the method may lose convergence. In the weak $U$ region, the FLEX can reproduce the Fermi surface obtained in CDMFT for $U\le$ 3.0 (see SM.5). However, FLEX fails to see a Fermi surface topology similar to Fig.~\ref{fig:CDMFT}{\bf k} obtained in CDMFT at $U=3.6$ eV. This indicates that FLEX can not describe the strong interlayer correlation which plays a crucial role in the electronic structure of the bilayer nickelate thin films.

Next, we discuss the comparison between strained thin films and pressurized bulks.
The coupling between interlayer $d_{z^2}$ orbitals ($t_\perp^z/J_\perp^z$) are believed to be crucial in the pressurized bulks\cite{Liu_2023_sPlusMinus,sakakibara2024,liao_2024}.  However, the thin films have a much longer interlayer distance than the that in the pressurized bulks, which leads to much weaker interlayer couplings, say about $30\%$ lower for $t^z_{\perp}$ and $50\%$ lower for $J^z_{\perp}$ (since it proportion to $(t^z_{\perp})^2$).  
It would be a challenge for the theories strongly depending on $t^z_{\perp}$ and/or $J^z_{\perp}$ to explain the same order-of-magnitude onset ${T_\mathrm{C}}$ of the two systems.
In contrast, although DFT calculations suggest a distinct Fermi surface topology for the pressurized samples, the CDMFT-derived Fermi surface (which incorporates correlation effects and aligns with ARPES data) closely resembles that of the high-pressure case. Therefore, it would be much easier to understand why the thin films have same order-of-magnitude onset ${T_\mathrm{C}}$ with a much smaller $t^z_{\perp}$ given that the correlated effect has been included correctly.

At last, we would like to emphasize the quantitative effectiveness of the DFT+cRPA+CDMFT approach in the study of unconventional superconductors with weak to intermediate correlation effects. Simply with lattice parameter inputs from experiments, the {\it ab initio} calculations without parameter fine-tuning match ARPES results quantitatively. The following RPA or similar techniques may provide a qualitatively accurate description of the pairing symmetry on top of the precise correlated electronic structures. Therefore, this approach is feasible for studying superconductivity in weakly and moderately correlated systems within first-principles calculations. Its wide applicability in various materials may shift the paradigm of searching for unconventional superconductors.

\begin{methods}

\hspace*{2em}{\bf STEM experiment.} All electron microscopy analyses were conducted using a FEI Themis Z transmission electron microscope operated at 200 kV. This microscope is equipped with a Cs Probe Corrector (DCOR) and a high-brightness field-emission gun (X-FEG) with a monochromator to enhance resolution and contrast. For scanning transmission electron microscopy (STEM) high-angle annular dark field (HAADF) and annular bright field (ABF) imaging, the inner and outer acquisition angles ($\beta$1 and $\beta$2) were set to 90 and 200 mrad for HAADF images, and 12 and 23 mrad for ABF images, respectively. The convergence angle was maintained at 25 mrad. The beam current was adjusted to approximately 40 pA for HAADF and ABF imaging. Cross-sectional STEM specimens were prepared using a FEI Helios G4 HX dual-beam focused ion beam/scanning electron microscope (FIB/SEM) system. The simulation of ABF images was performed using the multi-slice method implemented in the QSTEM software. The simulation parameters were carefully set to match those used during the electron microscopy characterization, ensuring that the simulated images accurately reflect the experimental conditions.

{\bf Ab initio multi-orbital Hubbard model.} 
We construct the realistic multi-orbital Hubbard model for the half-UC thin film, including the non-interacting tight-binding (TB) model and the onsite interaction term 
\begin{equation}
H=\sum_{{\bf k \sigma}}\Psi_{{\bf k \sigma}}^{\dagger}H^{0}({\bf k})\Psi_{\bf k \sigma}+\sum_{i\tau}H_{i\tau}^{\mathrm{int}}.
\label{eq:H_full}
\end{equation}
We use $i$ to enumerate the unit cells, $\tau$ the layer indices ($A$ for the bottom layer, $B$ for the top layer), $\gamma$ the orbital indices ($x$ for $d_{x^2-y^2}$ orbital and $z$ for $d_{z^2}$ orbital), $\sigma$ the spin states $\uparrow$ and $\downarrow$, and $\Psi_{\bf k \sigma}=\left(d_{\bf k Ax\sigma},d_{\bf k Az\sigma},d_{\bf k Bx\sigma},d_{\bf k Bz\sigma}\right)^{T}$.

The low-energy bands' TB model is obtained from the maximally localized Wannier functions constructed by the Wannier90 package\cite{wannier90,Pizzi_2020}. The TB model Hamiltonian in ${\bf k}$-space has the same matrix structure as in Ref.~\cite{YaoPRL2023}, reading 
\begin{align}
H^0({\bf k}) &=\left(\begin{array}{cc}
H^0_{A}({\bf k}) & H^0_{AB}({\bf k})\\
H^0_{AB}({\bf k}) & H^0_{A}({\bf k})
\end{array}\right), &
H^0_{A}({\bf k})&= \left(\begin{array}{cc}
T_{{\bf k}}^{x} & V_{{\bf k}}\\
V_{{\bf k}} & T_{{\bf k}}^{z}
\end{array}\right),& H^0_{AB}({\bf k})&=\left(\begin{array}{cc}
T_{{\bf k}}^{\prime z} & V_{{\bf k}}^{\prime}\\
V_{{\bf k}}^{\prime} & T_{{\bf k}}^{\prime z}
\end{array}\right).\label{eq:tb}
\end{align}
with 
\begin{align*}
T_{{\bf k}}^{x/z}&=2t_{1}^{x/z}(\cos k_{x}+\cos k_{y})+2t_{4}^{x/z}(\cos2k_{x}+\cos2k_{y})\\&+2t_{5}^{x/z}(\cos3k_{x}+\cos3k_{y})+4t_{2}^{x/z}\cos k_{x}\cos k_{y}+\epsilon_{x/z},\\
V_{{\bf k}}&=2t_{3}^{xz}(\cos k_{x}-\cos k_{y})+2t_{5}^{xz}(\cos2k_{x}-\cos2k_{y}),\\
T_{{\bf k}}^{\prime z}&=t_{\perp}^{x/z}+2t_{3}^{x/z}(\cos k_{x}+\cos k_{y}),\\
V_{{\bf k}}^{\prime}&=2t_{4}^{xz}(\cos k_{x}-\cos k_{y}).
\end{align*}
Here, $T^{x/z}_{\bf k}$ ($T^{\prime x/z}_{\bf k}$) is the intra-layer (inter-layer) intra-orbital hopping, and $V_{{\bf k}}$
($V_{{\bf k}}^{\prime}$) is the intra-layer (inter-layer) inter-orbital hopping. The hopping parameters are summarized in Table \ref{tab:tb_paras}. 

The interacting part of Eq.~\ref{eq:H_full} reads $H^{\mathrm{int}}=\sum_{i\tau} H^{\mathrm{int}}_{i\tau}$ with $H^{int}_{i\tau}$ the Kanamori type two-orbital onsite interaction  
\begin{align}
H_{\mathrm{i\tau}}^{\mathrm{int}}&=\sum_{\gamma}U_{\gamma} n_{i\tau\gamma\uparrow} n_{i\tau\gamma\downarrow}+\sum_{\gamma<\gamma',\sigma\sigma^{\prime}}(U^{\prime}-\delta_{\sigma\sigma^{\prime}}J)n_{i\tau\gamma\sigma}n_{i\tau\gamma'\sigma'}\nonumber \\
&-J\sum_{\gamma<\gamma'}(c_{i\gamma\uparrow}^{\dagger}c_{i\gamma\downarrow}^{\dagger}c_{i\gamma'\downarrow}c_{i\gamma'\uparrow}+c_{i\gamma'\uparrow}^{\dagger}c_{i\gamma'\downarrow}^{\dagger}c_{i\gamma\downarrow}c_{i\gamma\uparrow}+\mathrm{h.c.})
\label{eq:Hloc}
\end{align}
$H^{int}_{i\tau}$ contains the intra-orbital ($U$), inter-orbital ($U^\prime=U-2J$) Coulomb repulsions, and the Hund coupling $J$. The second line in Eq.~\ref{eq:Hloc} shows the spin-flip and pair-hopping terms. The realistic values of $U$ and $J$ are obtained by the constrained random phase approximation (cRPA)\cite{cRPA2004}, which counts in the screening effect from the rest energy bands out of the low-energy $e_g$ subspace. The results obtained from cRPA are listed in Table \ref{tab:cRPA}.  Although there is slight anisotropy between $U_x$ and $U_z$, we take their average $\overline{U}=\sum_\gamma U_\gamma / 2 \approx$3.77 eV for both orbital and the Hund coupling J as $J_{\mathrm{cRPA}}$ = 0.56 eV, which is slightly different from those of the high pressure bulk crystal (3.79 and 0.61 eV)\cite{Christiansson_2023}. 
We also listed the inter-layer Coulomb repulsion $V$ ranging in$\sim 1.0-1.3$ eV in the thin film, roughly 30\% of $\overline{U}_{\mathrm{cRPA}}$. 
The crystal structure, TB model obtained from wannier90 with all long-range hopping integrals and the cRPA interaction tensor $U_{ijkl}$ have been open-sourced in \href{https://zenodo.org/records/14634336}{Zenodo}\cite{Yue_327_data}.

{\bf CDMFT calculation.} We apply the CDMFT\cite{Georges1996,Kotliar2006} method to study the correlated electronic structure of the half-UC thin film. In the bonding and anti-bonding basis of ${z_+}$ ${z_-}$ ${x_+}$ ${x_-}$, the cluster impurity problem is equivalent to an effective four-orbital single impurity model with diagonal crystal field splitting among these orbitals. The numerically exact hybridization-expansion continuous-time quantum Monte Carlo  (CT-HYB)\cite{Werner2006,Gull2011} is adopted to solve the quantum impurity model. To calculate the momentum-resolved spectral function, we adopt the maximum-entropy analytic continuation method to continue the self-engery from the Matsubara frequency axis to the real frequency axis\cite{Jarrell_1996_Bayesian}. 

{\bf CDMFT+RPA calculation.} The CDMFT+RPA method is a modified RPA approach in which the infinite sum of bubble diagrams, i.e., the random phase approximation (RPA), is performed upon the quasi-particle Hamiltonian from CDMFT. We assume that the effective interaction has the same form as bare interaction with strength only renormalized by the quasi-particle weight. The advantage of the CDMFT+RPA method is that small effective interaction actually corresponds to intermediate bare interaction. The RPA-renormalized spin and charge susceptibilities can investigate spin-density wave (SDW), charge-density wave (CDW), and pairing instabilities. There exists critical interaction strength, above which RPA results become unreliable. Near critical values, enhanced spin fluctuations can mediate an effective attraction between quasi-particles close to the Fermi surface and serve as glues for forming Cooper pairs. The effective attraction is treated by mean-field theory to derive the self-consistent SC gap equation. Near SC critical temperature $T_{c}$, the gap equation can be linearized and becomes a standard eigenvalue problem. The largest eigenvalue of the effective pairing interaction vertex matrix determines $T_{c}$, and corresponding eigenvectors give the distribution of gap functions, from which we can determine the pairing symmetry\cite{Scalapino_1986}.

\end{methods}

\begin{addendum}

\item[Acknowledgements] 
This work was supported by the National Key Research and Development Program of China (2024YFA1408101 and 2022YFA1403101), the Natural Science Foundation of China (1247041908, 12141402, 12404171, 92265112, 12374455, and 52388201), Shenzhen Municipal Funding Co-Construction Program Project (SZZX2401001 and SZZX2301004), Guangdong Provincial Quantum Science Strategic Initiative (GDZX2401004 and GDZX2201001), and Guangdong Provincial Key Laboratory of Advanced Thermoelectric Materials and Device Physics (2024B1212010001). Part of the calculations were carried out at the Center for Computational Science and Engineering at the Southern University of Science and Technology. Zhuoyu Chen also acknowledges the support of the Shenzhen Science and Technology Program (KQTD20240729102026004).

\item[Author contributions]

W.Q.C., C.M.Y., and Z.C. designed the research. C.M.Y. performed the DFT and  CDMFT calculations. J.J.M. performed the RPA calculations. H.H. conducted the STEM analysis with assistance from Q.Y., and J.L.. Y.C.H. performed the FLEX calculations.  P.L. and Y.L. provided ARPES data. G.Z. and W.L. provided samples. F.Y., H.S., Y.J.S., Q.K.X., and all other authors participated in discussions. W.Q.C., C.M.Y., J.J.M., Z.C., H.H., and Y.C.H. wrote the manuscript.

\item[Competing financial interests] The authors declare that they have no competing financial interests.

\end{addendum}

\newpage

\begin{table}
\centering
\caption{ |
{\bf The lattice parameters of La$_{2.85}$Pr$_{0.15}$Ni$_2$O$_7$ film on substrate SrLaAlO$_4$.} STEM measurements are calibrated by the lattice constants obtained from X-ray diffraction experiments.
The term "La-La intra." refers to the distance between La atoms in the c-axis direction within the perovskite layer, while "La-La inter." refers to the distance between adjacent La atoms in the c-axis direction between the perovskite bilayers. The abbreviation "sub." stands for "substrate". Fig. S2 shows the schematic indications of these spacings.
}
\begin{tabular}{ |>{\bfseries}m{3.5cm}|m{2.3cm}|m{2.3cm}|}
\hline
\ & XRD & STEM \\
\hline
a ({\AA}) & $3.754$ & - \\
b ({\AA}) & $3.754$ & - \\
c ({\AA}) & $20.819$ & -\\
\hline
Ni-O-Ni angle ($^\circ$) & - & $180\pm 5$   \\
\hline
Ni-Ni length ({\AA})  & - & $4.28\pm 0.05$   \\
\hline
La-La intra. ({\AA}) & - & $3.71\pm 0.05$ \\
\hline
La-La inter. ({\AA}) & - & $3.0\pm 0.05$ \\
\hline
\end{tabular}
\label{tab:doped_film_paras}
\end{table}

\begin{table}
\centering
\caption{ | 
{\bf The tight-binding parameters (in a unit of eV) for the half-UC thin film obtained from DFT+U}. The number in the brackets corresponds to that of the high-pressure bulk structure reported by Luo {\it et al.}\cite{YaoPRL2023}.
}
\begin{tabular}{|c|c|c|c|c|}
\hline
$t_{1}^{x}$ & $t_{1}^{z}$ & $t_{2}^{x}$ & $t_{2}^{z}$ & $t_{3}^{xz}$      \tabularnewline\hline 
-0.466 (-0.483) & -0.126 (-0.110) & 0.062 (0.069) & -0.016 (-0.017) & 0.229 (0.239) \tabularnewline \hline\hline
 $t_{\bot}^{x}$ & $t_{\bot}^{z}$ & $t_{4}^{xz}$ & $\epsilon^{x}$ & $\epsilon^{z}$  \tabularnewline \hline 
0.001 (0.005) & -0.439 (-0.635) & -0.032 (-0.034) &  0.870 (0.776) & 0.351 (0.409) \tabularnewline \hline\hline
 $t_{4}^{x}$ & $t_{4}^{z}$ & $t_{5}^{x}$ & $t_{5}^{z}$ & $t_{5}^{xz}$ \tabularnewline \hline 
 -0.064 & -0.014 &  -0.015 & -0.003 & 0.026 \tabularnewline \hline\hline
 $t_{3}^{x}$ & $t_{3}^{z}$ &             &             &              \tabularnewline \hline 
0.033 & -0.001 & & & \tabularnewline 
\hline 
\end{tabular}
\label{tab:tb_paras}
\end{table}

\begin{table}
\centering
\caption{ 
{\bf Interaction parameters}
The static intra-layer on-site Coulomb repulsion  ($U$) and inter-layer non-local Column interaction $V$ (in a unit of eV) calculated from cRPA  $\overline{U}$ is the orbital averaged value.}
\begin{tabular}{ccccccc}
\hline \hline 
$U^x$ & $U^z$ & $\overline{U}$ & $J$ & $V_{\perp}^x$ & $V_{\perp}^z$ & $V_{\perp}^{x,z}$ \tabularnewline\hline
4.03 & 3.51 & 3.77 &  0.56 & 1.04 & 1.31 & 1.13 \tabularnewline
\hline \hline
\end{tabular}
\label{tab:cRPA}
\end{table}

\newcommand{\beginsupplement}{%
        \setcounter{table}{0}
        \renewcommand{\thetable}{S\arabic{table}}%
        \setcounter{figure}{0}
        \renewcommand{\thefigure}{S\arabic{figure}}%
}

\title{\textbf{Supplementary Information }}

\maketitle

\beginsupplement

\section*{SM1. Additional structural information}

The lattice structure parameters were taken from a superconducting  La$_{2.85}$Pr$_{0.15}$Ni$_2$O$_7$ film and a non-superconducting La$_3$Ni$_2$O$_7$ film grown on SrLaAlO$_4$ substrates\cite{Zhou_2024_AmbientSC}. Fig. S1 shows the annular bright-field (ABF) images of the cross-section of La$_{2.85}$Pr$_{0.15}$Ni$_2$O$_7$ sample, projected along [100]$_{pc}$, [010]$_{pc}$ and [110]$_{pc}$. We analyzed the positions of oxygen atoms in these three projections and selected part of the data to show in Fig. 1, which reflects that the La$_{2.85}$Pr$_{0.15}$Ni$_2$O$_7$ thin film has almost no oxygen octahedral rotation under compressive epitaxial strain. The atomic spacing used for modeling is obtained by measuring directly on High-angle Annular Dark Field (HAADF) images with calibration from XRD lattice constants. Fig. S2 identifies several key atomic spacing values of La$_{2.85}$Pr$_{0.15}$Ni$_2$O$_7$ and La$_3$Ni$_2$O$_7$ films, which are also listed in Tables 1 and Tables S1. At the same time, considering the error of the electron microscope in the measurement, the listed values show only two significant digits after the decimal point.

\section*{SM2. Ab initio Calculations}
 We build the crystal structure of the half-UC ultra-thin film of La$_3$Ni$_2$O$_7$ according to the structural parameters (see Table 1 in the main text) of the ultra-thin film of La$_{2.85}$Pr$_{0.15}$Ni$_2$O$_7$. To be specific, we choose the lattice constant $a=b=3.7544${\AA}, inter-layer La-La distance 3.705{\AA}, inter-layer Ni-Ni distance 4.28{\AA}. The middle La layer is placed in the middle of the Ni-bilayer. Furthermore, the inter-layer Ni-O-Ni and La-O-La angles are  180$^\circ$ as the ultra-thin film has $C_4$ rotation symmetry. As for the $c$ axis, we set $|c|$=40 {\AA} to make the length of the vacuum $\sim$ 30{\AA} long enough to simulate thin film in the periodic structure. The positions of oxygen are placed according to its high-pressure structure. The {\it ab initio} density-function theory (DFT) calculations, including structure relaxation and band structure, are performed using the VASP package\cite{vasp_ref1,vasp_ref2,vasp_ref3}, in which we use the Perdew-Burke-Ernzerhof (PBE) exchange-correlation functional\cite{PBE_ref}. As the ions are heavier in mass and the strain from the substrate confines their positions, we relax only the oxygen atoms in the built structure while keeping the La and Ni atoms fixed. In the DFT calculations, we set the energy cutoff for the plane-wave basis 550 eV, k-mesh grid $11\times 11 \time 1$. The atomic positions of oxygen are relaxed until the norms of all the forces are smaller than 0.001 eV/{\AA}.

 We adopt the virtual crystal approximation (VCA) to simulate the Sr-doping effect. As for where to dope Sr, there are several different methods. Here, we tried two different kinds of doping: doping Sr only to the middle La layer and doping Sr uniformly to all La layers. 
 In the main text, the band structure and tight-binding model correspond to the results of doping Sr only to the middle La layer. The following shows that the low-energy bands are not sensitive to these two doping methods. 
 First, we confirm that the low-energy bands of Ni-$e_g$ orbitals are nearly rigid, especially the bands near Fermi energy, as shown by Fig.~\ref{fig:SM_Fig1_DFT_compare}{\bf a}. 
 Second, we confirm that the low-energy bands Ni-$e_g$ orbitals are insensitive to how we dope Sr to La when the average hole doping level per Ni-layer is the same, as shown by Fig.~\ref{fig:SM_Fig1_DFT_compare}{\bf b} (doping Sr only to the middle La layer) and  {\bf c} (doping Sr uniformly to all La layer). We also directly compare the DFT+U bands between two different methods of Sr-doping in Fig.~\ref{fig:SM_Fig1_DFT_compare}{\bf d}, confirming that the low-energy bands of Ni-$e_g$ orbitals are almost identical even though the high energy bands are different. The relaxed atomic positions corresponding to doping Sr only in the middle La layer are tabulated in Table~\ref{tab:struct_parameters}. In both Sr-doping methods, the low-energy bands are insensitive to $U$ as shown in Fig.~\ref{fig:SM_Fig1_DFT_compare}{\bf b-c}. Here, the effect of $U$ is to push downward the already full-filled $t_{2g}$ bands. 

 As the low-energy Ni-$e_g$ bands are almost identical to each other, no matter where Sr is doped and whether or not $U$ is added, these TB parameters calculated with DFT+$U$ can also be safely regarded as that of the DFT bands, as usually used in the literature. The directions of all hopping integrals are schematically demonstrated in Fig.\ref{fig:TBmodel}. Besides two key differences between the TB model of the half-UC ultra-thin film and that of high-pressure bulk, we introduce six additional long-range hopping parameters. For example, we have the inter-layer next nearest neighbor (NNN) hopping between the $x$-orbitals $t^x_3$=0.0332 eV, the intra-layer NNN hopping between $x$-orbitals $t^x_4$=-0.0639 eV. There is also inter-layer inter-orbital NNNN hopping $t^{xz}_5$=0.0255 eV. 

\section*{SM3. Doping and Temperature Dependence of CDMFT results}  

Fig.~\ref{fig:SM_CDMFT_U3p6_doping} shows the doping dependence of $A({\bf k},\omega)$ and FS at $U=3.6$ eV and indicated fillings. As one decreases the filling $n=1.4$ to $n=1.2$ (equivalent to increasing hole doping from 0.1 hole to 0.3 hole per Ni-$e_g$ orbitals), one can see that the $\gamma$  ($\delta$) band moves upward (downward), which can be explained by a reduced effective level splitting between $z_+$ and $z_-$ orbitals. The Fermi surface at $n=1.2$ (Fig.~\ref{fig:SM_CDMFT_U3p6_doping}{\bf f}) resembles the one seen in DFT (Fig.~2{\bf c}) at $n=1.33$. The reason is that the inter-orbital correlation strength decreases when the filling per orbital is reduced. As a result, the $\gamma$ hole ($\delta$ electron) pocket gradually appears and expands as $n$ reduces. 

Fig.~\ref{fig:SM_CDMFT_U3p6_n_1p3_T_dependence} shows the temperature dependence of $A({\bf k},\omega)$ and FS at $U=3.6$, which changes only qualitatively as one decreases of temperature from $T$=200 K to $T$=50 K. The main difference is that the low-energy electrons become more coherent at lower $T$, leading to sharper low-energy bands and sharper Fermi surfaces. 

\section*{SM4. CDMFT+RPA} 

CDMFT+RPA is based on the quasi-particle Hamiltonian from CDMFT . As shown by our CDMFT calculation, the Fermi surface topology can be greatly affected at larger $U$>3 eV. As the pairing is mainly controlled by the low-energy excitations, a better start point for RPA is to use the quasi-particle Hamiltonian obtained from CDMFT. We call such treatment CDMFT+RPA. The   quasi-particle Hamiltonian $\tilde{H}_{{\bf k}}^{\mathrm{QP}}$ is obtained from the quasi-particle approximation of self-energy 
\begin{equation}
\tilde{\mathbf{\Sigma}}^{\mathrm{QP}}(\omega)\approx\mathrm{Re}\tilde{{\bf \Sigma}}(i0^{+})+(\mathbb{\mathbf{I}}_{4}-\tilde{\mathbb{\mathbf{Z}}}_{4}^{-1})(\omega+i\eta)
\end{equation}
Here, we do not distinguish spin as they are degenerate (dimension 4 for each spin).  $\tilde{\ }$ means all the quantities are in the bonding and anti-bonding basis. In this basis, the zero-frequency self-energy $\tilde{{\bf \Sigma}}(i0^{+})$ and quasi-particle spectral weight $\tilde{\mathbb{\mathbf{Z}}}$ obtained from CDMFT are both diagonal. The quasi-particle Green's function reads
\begin{align}
\tilde{\mathbf{G}}^{\mathrm{QP}}({\bf k},\omega)&=\left[(\omega+i\eta+\mu)\mathbf{I}_{4}-\tilde{H}_{{\bf k}}-\tilde{\mathbf{\Sigma}}^{\mathrm{QP}}(\omega)\right]^{-1}  
\end{align}
The corresponding quasi-particle Hamiltonian is 
\begin{align}
\tilde{H}_{{\bf k}}^{\mathrm{QP}}=\tilde{\mathbb{\mathbf{Z}}}^{1/2}[\tilde{H}_{{\bf k}}-\mu\mathbf{I}_{4}+\mathrm{Re}\tilde{{\bf \Sigma}}(i0^{+})]\tilde{\mathbb{\mathbf{Z}}}^{1/2}
\end{align}

FIG.~\ref{fig:fig_SM_Q1} illustrates that the correspondence between $\mathbf{Q}_{1}$ in the distribution of RPA-renormalized spin susceptibility and FS nesting still holds for different parameters. As the shape of FS changes for different fillings and bare interactions in DMFT, the wave vector $\mathbf{Q}_{1}$ also varies accordingly. FIG.~\ref{fig:fig_SM_Q1}{\bf b,d,f,h} demonstrate that the gap functions on FS patches connected by $\mathbf{Q}_{1}$ possess opposite signs, which is similiar to the $d_{x^2-y^2}$-wave pairing in the cuprates and the $s^{\pm}$-wave pairing in the iron-based superconductors.
    
FIG.~\ref{fig:fig_SM_lam} presents the dependence of $\lambda$ on $\ensuremath{U_{\textrm{eff}}}$ for $U=3.77$ eV, $U=3.6$ eV, and $U=3$ eV. The leading pairing symmetry is always the $s^{\pm}$ wave pairing before RPA-renormalized spin susceptibility diverges. For small bare interaction in FIG.~\ref{fig:fig_SM_lam}{\bf c}, there is obvious competition between $s^{\pm}$-wave and $d_{xy}$-wave pairings. With stronger bare interaction, only the $s^{\pm}$-wave pairing instability is leading as shown in FIG.~\ref{fig:fig_SM_lam}{\bf a}. Both results further support the main conclusion that the most possible pairing symmetry is $s^{\pm}$-wave pairing for realistic parameters in the main text.
    
\section*{SM5. FS in Fluctuation-exchange approximation}
Fluctuation-exchange (FLEX) approximation\cite{bickers1989} has been developed to study the pairing instability of Hubbard models. FLEX is by nature a weak-coupling approach and it breakdowns when $U$ is stronger. Compared with normal RPA, FLEX can reach slightly higher correlation strength as the self-energy is incorporated. Here, we apply the FLEX calculation to the same multi-orbital Hubbard model of the ultra-thin film using a modified FLEX package developed by Witt {\it et al.}\cite{witt2021}. The Fermi surfaces at filling $n=1.3$ are shown in Fig.~\ref{fig:FLEX-FS}{\bf a-b}. It turns out the $\delta$-pocket remains at $\Gamma$ point when $U$ is increased from $U=1$ to $U=1.8$ above which FLEX breaks down. While the FS of FLEX at the same filling $n=1.3$ is similar to that given by CDMFT at $U\le 3$ (Fig.~3{\bf i-j}), FLEX fails to reproduce FS as in Fig.~3{\bf l} in the safe radius of $U$. 
Note that the largest $U$ that FLEX can reach depends on the filling and density of states. We confirm that our FLEX calculation can reach $U\le 3$ for the TB model of high-pressure bulk crystal, which has a less flat $\gamma$-band than the ultra-thin film.


\begin{table}
\centering
\caption{ 
| The lattice parameters of La$_3$Ni$_2$O$_7$ film and substrate SrLaAlO$_4$. STEM measurements are calibrated by the lattice constants obtained from XRD experiments. The term "La-La intra" refers to the distance between La atoms in the c-axis direction in the perovskite layer. The term "La-La inter" refers to the distance between adjacent La atoms in the c-axis direction between the perovskite bilayers. The "cif" column lists the parameters from standard ".cif" file. All lattice parameters are represented numerically using pseudocube cells.
}
\begin{tabular}{ |>{\bfseries}m{3.5cm}|m{1.5cm}|m{1.5cm}|m{2cm}|m{1.5cm}|m{2cm}|m{1.5cm}|}
\hline
 & \multicolumn{3}{c|}{\textbf{SrLaAlO$_4$}} & \multicolumn{3}{c|}{\textbf{La$_3$Ni$_2$O$_7$}} \\
\hline
 & \textbf{cif} & \textbf{XRD} & \textbf{STEM} & \textbf{XRD} & \textbf{STEM} & \textbf{cif}  \\
\hline
a & $3.7544$  & $3.754$ & $3.75$ & $3.754$ & - & $3.833$  \\
b & $3.7544$  & $3.754$ & $3.75$ & $3.754$ & - & $3.833$  \\
c & $12.649$ & $12.630$ & $12.64$ & $20.758$ & - & $20.518$ \\
\hline
Ni-O-Ni angle ($^\circ$) & - & - & - & - & $180\pm 5$ & -\\
\hline
Ni-Ni length & - & - & - & - & $4.04\pm 0.05$ &  - \\
\hline
La-La intra & - & - & $3.60\pm 0.05$ & - & $3.68\pm 0.05$ & -\\
\hline
La-La inter & - & - & $2.75\pm 0.05$  & - & $3.09\pm 0.05$ & - \\
\hline
\end{tabular}
\end{table}

\begin{table}
\centering
\caption{ | Lattice constants and fractional coordinates of non-equivalent atomic positions for the crystal structure of the half-UC La$_3$Ni$_2$O$_7$ ultra-thin film. 
The Cartesian coordinate of each atom is given by $f_1 {\bf a}+ f_2 {\bf b}+ f_3 {\bf c}$.
In the VCA calculation to simulate Sr-doping, we either dope Sr {\it only} to the middle La layer with the ratio La:Sr=2:1 in this layer or dope Sr to {\it all} La layers with the ratio La:Sr=8:1. 
The crystal structure shown in Fig.~2{\bf a} can be reproduced with the space group P4/mmm (No.~123; the actual 2D space group is p4/mm).  {\it Make sure that $|c|$ is set 40 {\AA} to get correct spacing between atoms.}
}
\begin{tabular}{ccccc}
\hline \hline 
lattice constants & & & \tabularnewline \hline 
${\bf a}$({\AA}) & 3.7544 & 0 & 0 \tabularnewline \hline 
${\bf b}$({\AA}) & 0 & 3.7544 & 0 \tabularnewline \hline 
${\bf c}$({\AA}) & 0 & 0 & 40 \tabularnewline \hline 
frac. coord.    &  $f_1$ & $f_2$ & $f_3$ \tabularnewline\hline
La1  &  0.00000 &  0.00000 &  0.40738 \tabularnewline\hline  
 La2  &  0.00000 &  0.00000 &  0.50000 \tabularnewline\hline  
 Ni1  &  0.50000 &  0.50000 &  0.44651 \tabularnewline\hline  
  O1  &  0.50000 &  0.50000 &  0.39523 \tabularnewline\hline  
  O2  &  0.00000 &  0.50000 &  0.44793 \tabularnewline\hline  
  O3  &  0.50000 &  0.50000 &  0.50000 \tabularnewline\hline  
\hline 
\end{tabular}
\label{tab:struct_parameters}
\end{table}

\begin{figure*}[htp]
\centerline{\epsfig{figure=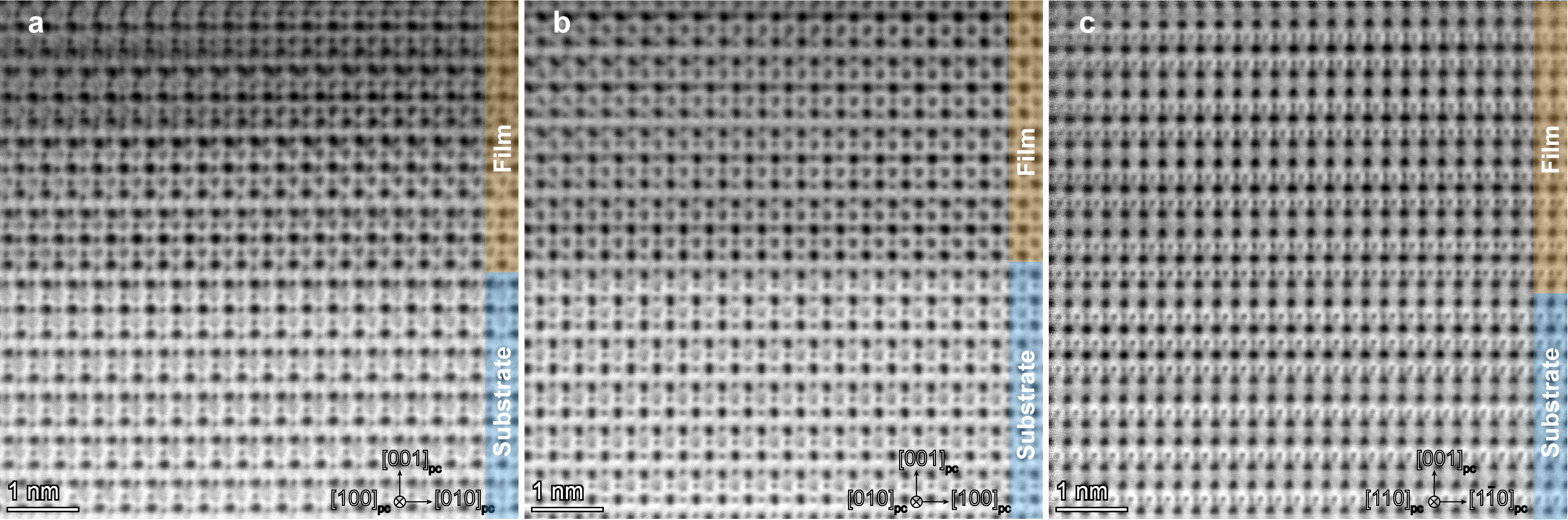,width=0.8\paperwidth}}
\caption{|
{\bf Annular bright-field (ABF) images of the cross-section of La$_{2.85}$Pr$_{0.15}$Ni$_2$O$_7$ film.} {\bf a-c}, The large-scale ABF images, projected along [100]$_{p}$, [010]$_{pc}$ and [110]$_{pc}$, respectively, delineate the boundaries of the film and the substrate. 	
}
\label{fig:Oxygen_rotaion_detail_SI}
\end{figure*}

\begin{figure*}[htp]
\centerline{\epsfig{figure=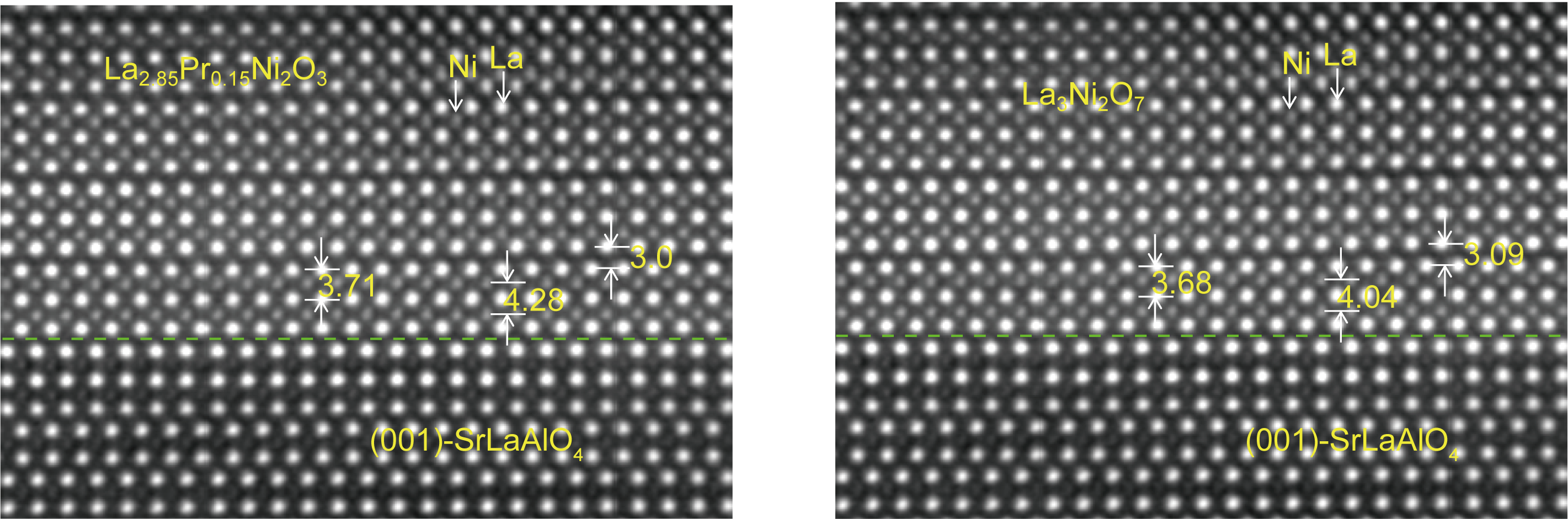,width=0.8\paperwidth}}
\caption{|{\bf High angle Annular Dark Field (HAADF) images with atomic spacings indicated. } {\bf a}, Atomic spacing of La$_{2.85}$Pr$_{0.15}$Ni$_2$O$_7$ film on SrLaAlO$_4$ substrate. {\bf b}, Atomic spacing of the La$_{3}$Ni$_2$O$_7$ film on SrLaAlO$_4$ substrate. The atom spacings in the substrates and the films are calibrated using the XRD measurements. 
}
\label{fig:stem_lattice_constants}
\end{figure*}

\begin{figure*}[htp]
\centerline{\epsfig{figure=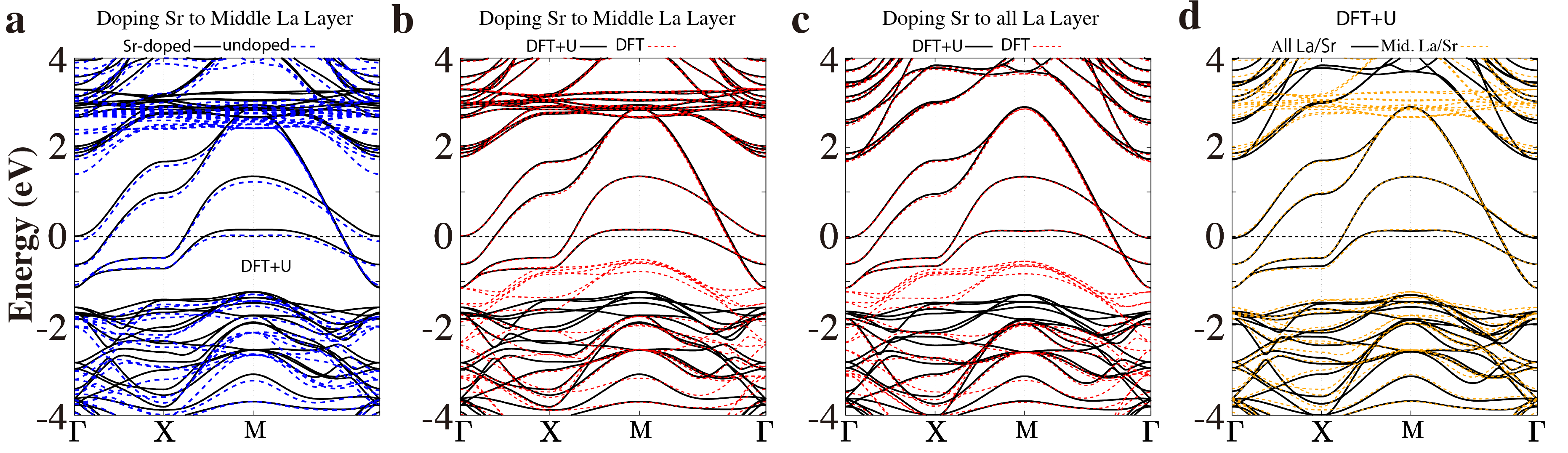,width=0.8\paperwidth}}
\caption{| {\bf Band structures of half-UC ultra-thin film calculated under several indicated conditions.} 
All the bands (black line in {\bf a} and all lines in {\bf b-d}) correspond to a filling of $n\approx1.33$ (1/6 hole) in Ni-$e_g$ orbitals, except the blue dashed line in {\bf a} corresponding to Sr un-doped system. $U=5$ and $J=$1 eV are used 1in all the DFT+$U$ calculations. 
{\bf a} Comparison between DFT+U bands of Sr-doped and of Sr-undoped system when Sr is doped only to the middle layer of La in the ultra-thin film with the composition ratio La:Sr=2:1 in the middle layer (8:1 in total when counting all La over Sr). 
{\bf b} Comparison between DFT+U bands (black solid line) and DFT bands (red dashed line) when Sr is doped only to the middle La layer. 
{\bf c} Comparison between DFT+U bands (black solid line) and DFT bands (red dashed line) when Sr is uniformly doped to all La layers with the composition ratio 8:1. 
{\bf d} Comparison between DFT+U bands of uniform Sr doping (black solid line) and of doping Sr only the middle layer (orange dashed line). 
}
\label{fig:SM_Fig1_DFT_compare}
\end{figure*}

\begin{figure*}[htp]
\centerline{\epsfig{figure=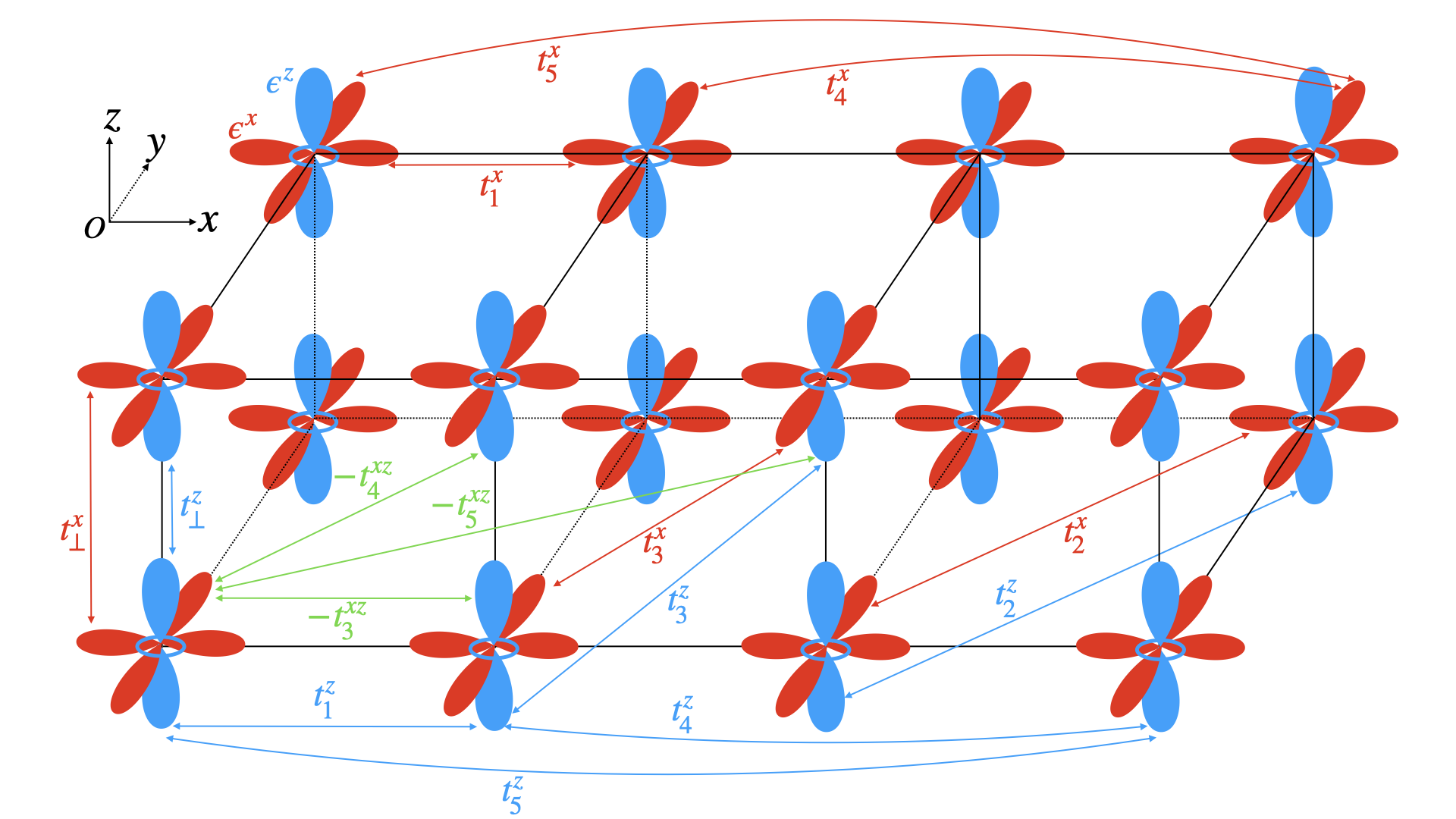,width=0.5\paperwidth}}
\caption{| \textbf{ Schematic illustration for all parameters of the tight-binding model.}}
\label{fig:TBmodel}
\end{figure*}

\begin{figure*}[htp]
\centerline{\epsfig{figure=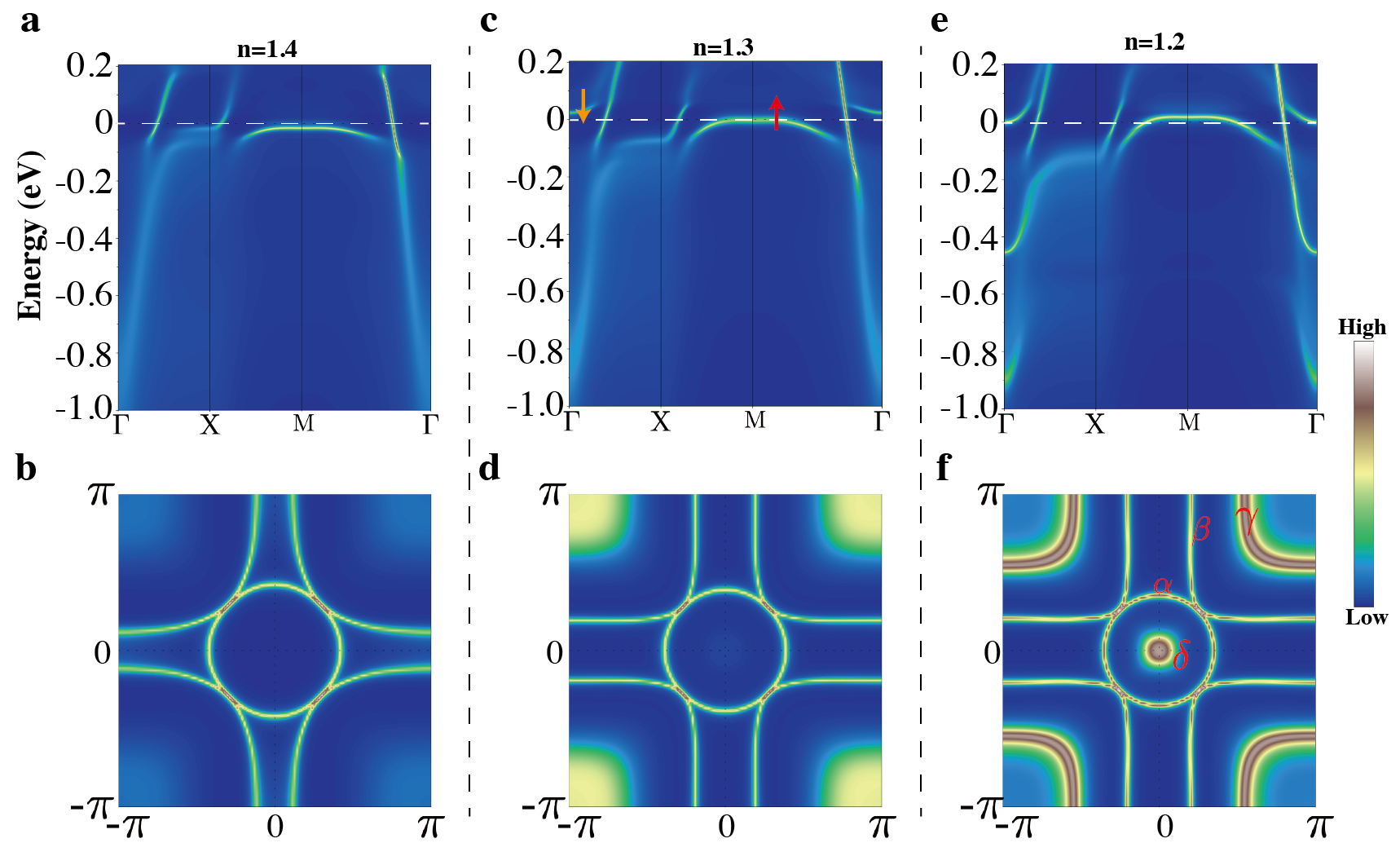,width=0.8\paperwidth}}
\caption{| {\bf Filling $n$-dependence of $A({\bf k},\omega)$ (top panels) and FS (lower panels) calculated at $U=$3.6 eV, $J=$0.56}. 
In all panels,  $U=3.6$ eV, $J$=0.56 eV, $T\sim$106K. 
}
\label{fig:SM_CDMFT_U3p6_doping}
\end{figure*}

\begin{figure*}[htp]
\centerline{\epsfig{figure=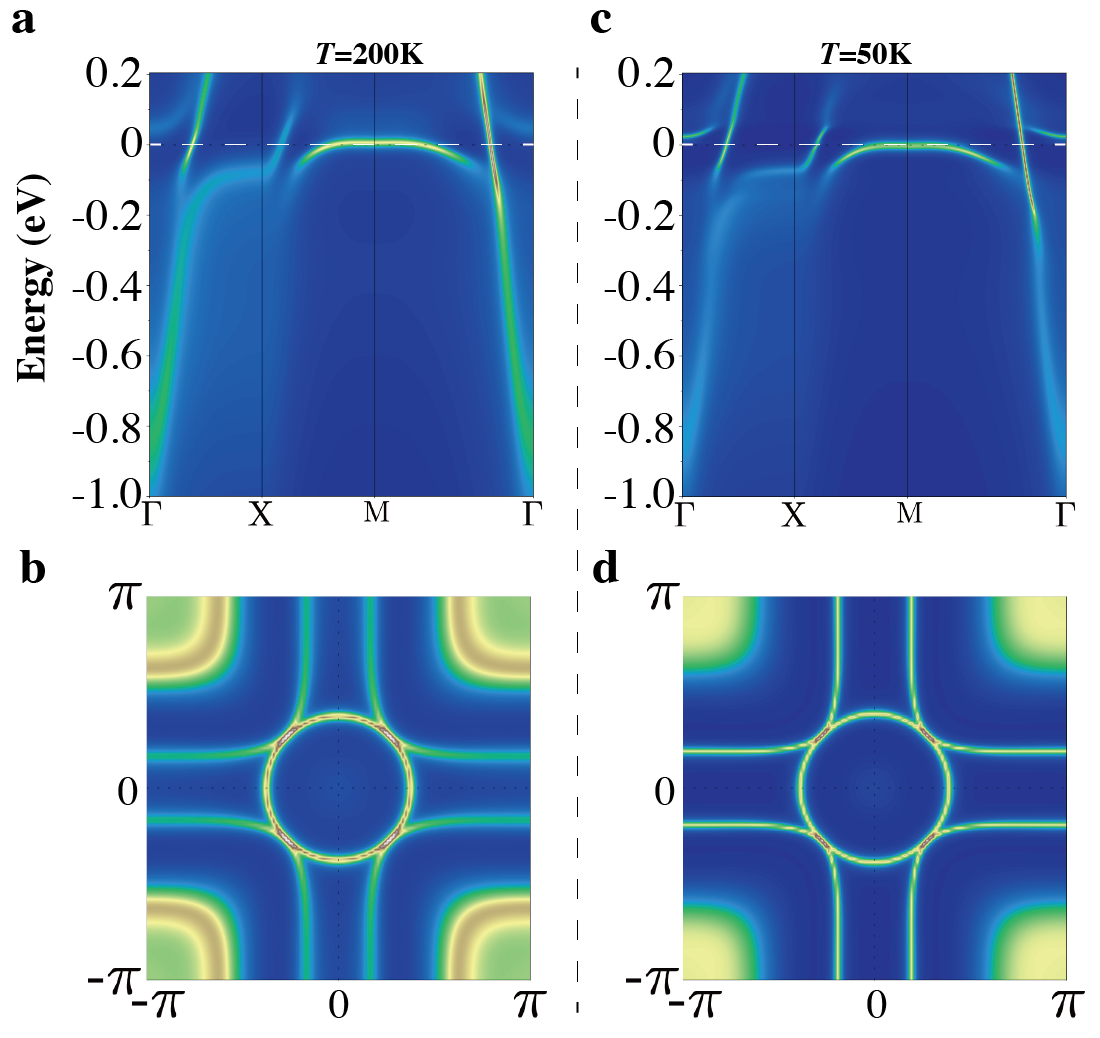,width=0.5\paperwidth}}
\caption{| {\bf Temperature dependence of $A({\bf k},\omega)$ (top panels) and FS (lower panels) calculated in CDMFT at $U=3.6$ eV, $J$=0.56 eV, $n$=1.3.} {\bf a-b} $T$=200 K; {\bf c-d} $T$=50K.
}
\label{fig:SM_CDMFT_U3p6_n_1p3_T_dependence}.
\end{figure*}

\begin{figure*}
\centerline{\epsfig{figure=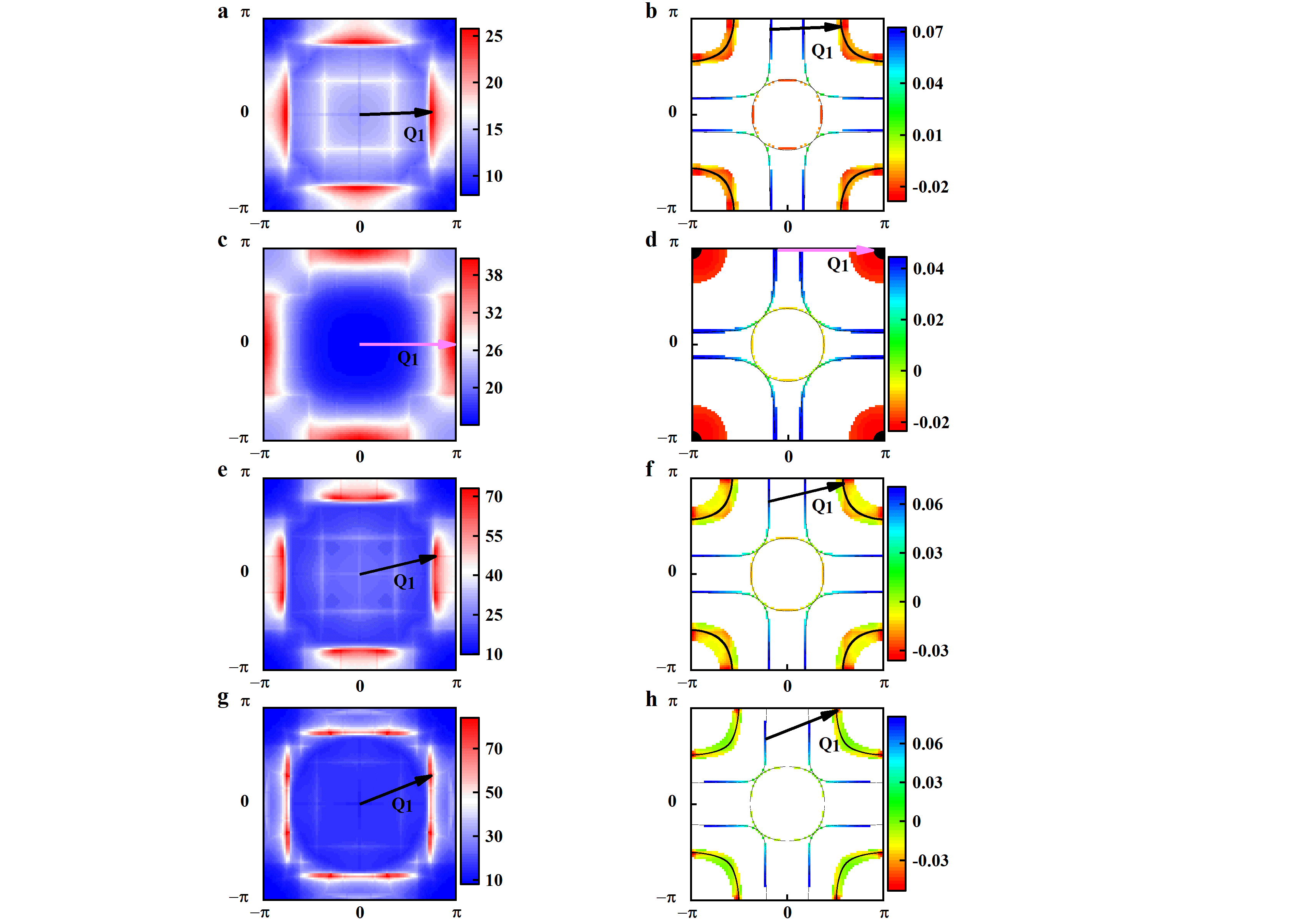,width=1.2\paperwidth}}
\caption{| {\bf The spin susceptibilities and corresponding gap functions on the FS.} {\bf a, b} $n=1.25$, $U=3.6$ eV and $\ensuremath{U_{\textrm{eff}}}=0.17$eV. {\bf c, d} $n=1.3$, $U=3.77$ eV and $\ensuremath{U_{\textrm{eff}}}=0.14$eV. {\bf e, f} $n=1.3$, $U=3.6$ eV and $\ensuremath{U_{\textrm{eff}}}=0.205$eV. {\bf g, h} $n=1.3$, $U=3$ eV and $\ensuremath{U_{\textrm{eff}}}=0.312$eV. 
}
\label{fig:fig_SM_Q1}
\end{figure*}

\begin{figure*}[htp]
\centerline{\epsfig{figure=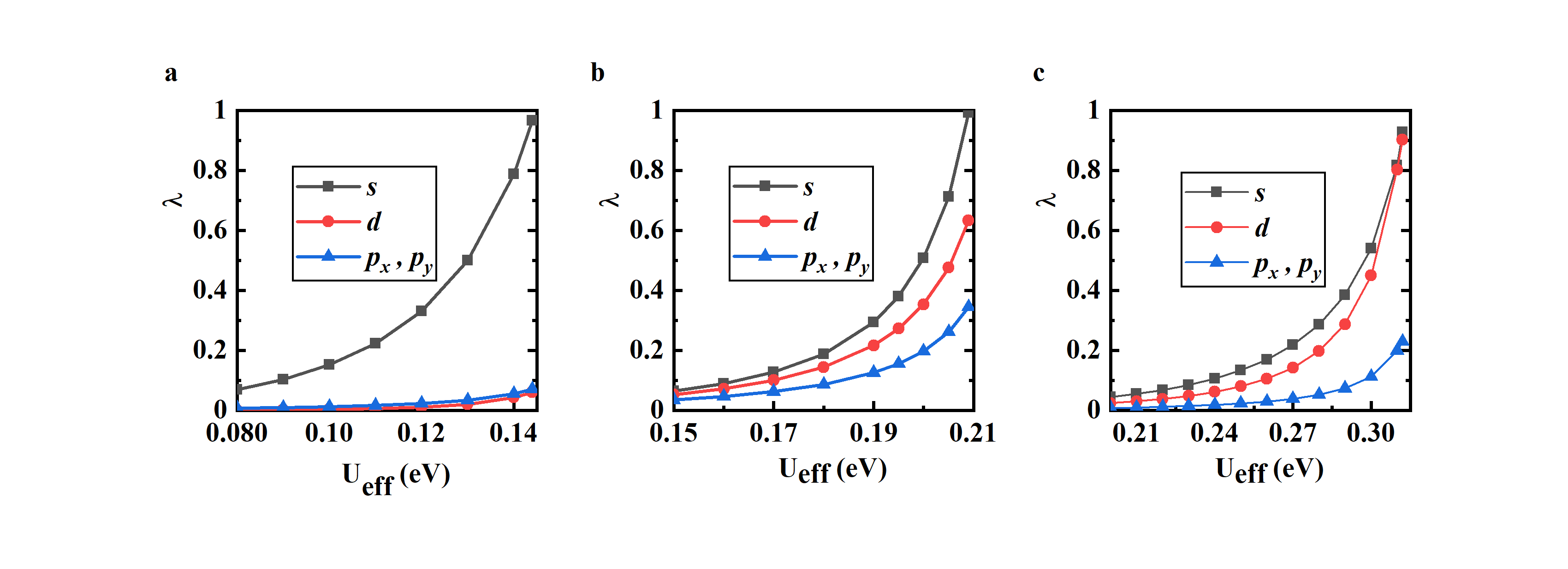,width=1.0\paperwidth}}
\caption{| {\bf The dependence of $\lambda$ on $\ensuremath{U_{\textrm{eff}}}$}. {\bf a} $U=3.77$ eV. {\bf b} $U=3.6$ eV. {\bf c} $U=3$ eV. The filling is $n=1.3$.  
}
\label{fig:fig_SM_lam}
\end{figure*}

\begin{figure*}[htp]
\centerline{\epsfig{figure=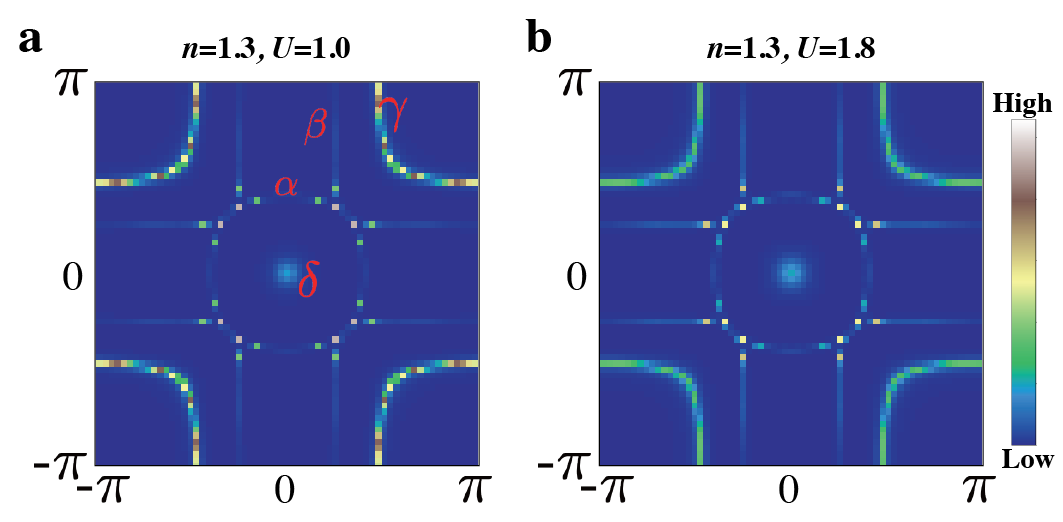,width=0.6\paperwidth}}
\caption{| {\bf Fermi surfaces calculated within FLEX at indicated $U$ and $n$}.
The temperature is $T=$0.001 eV ($\sim$ 11.6K), $J=U/6$,  
}
\label{fig:FLEX-FS}.
\end{figure*}



\section*{References}
\begin{thebibliography}{99}
\bibitem{Wang327_Nature2023} H. Sun, M. Huo, X. Hu, J. Li, Z. Liu, Y. Han, L. Tang, Z. Mao, P. Yang, B. Wang, J. Cheng, D.-X. Yao, G.-M. Zhang, and M. Wang, \textit{Signatures of superconductivity near 80 K in a nickelate under high pressure}, Nature (London) 621, 493 (2023).
\bibitem{hou2023emergence} J. Hou, P.-T. Yang, Z.-Y. Liu, J.-Y. Li, P.-F. Shan, L. Ma, G. Wang, N.-N. Wang, H.-Z. Guo, J.-P. Sun, \textit{et al.}, \textit{Emergence of high-temperature superconducting phase in pressurized La$_3$Ni$_2$O$_7$ crystals}, Chin. Phys. Lett. \textbf{40}, 117302 (2023).
\bibitem{Zhang_2024_HighTc} Y. Zhang, D. Su, Y. Huang, Z. Shan, H. Sun, M. Huo, K. Ye, J. Zhang, Z. Yang, Y. Xu, Y. Su, R. Li, M. Smidman, M. Wang, L. Jiao, and H. Yuan,\textit{High-temperature superconductivity with zero resistance and strange-metal behaviour in La$_3$Ni$_2$O$_{7-\delta}$}, Nat. Phys. \textbf{20}, 1269--1273 (2024).
\bibitem{Xie_2024_MagneticExchange}T. Xie, M. Huo, X. Ni, F. Shen, X. Huang, H. Sun, H. C. Walker, D. Adroja, D. Yu, B. Shen, L. He, K. Cao, and M. Wang,\textit{Strong interlayer magnetic exchange coupling in La$_3$Ni$_2$O$_{7-\delta}$ revealed by inelastic neutron scattering},Sci. Bull. \textbf{69}, 3221--3227 (2024).
\bibitem{Li_2024_ElectronicCorrelation}Y. Li, X. Du, Y. Cao, C. Pei, M. Zhang, W. Zhao, K. Zhai, R. Xu, Z. Liu, Z. Li, J. Zhao, G. Li, Y. Qi, H. Guo, Y. Chen, and L. Yang,\textit{Electronic correlation and pseudogap-like behavior of high-temperature superconductor La$_3$Ni$_2$O$_7$}, Chin. Phys. Lett. \textbf{41}, 087402 (2024).
\bibitem{Wang_2024_La2PrNi2O7_Nature} N. Wang, G. Wang,X. Shen, J. Hou,J. Luo, X. Ma, H. Yang, L. Shi, J. Dou, J. Feng, J. Yang, Y. Shi, Z. Ren, H. Ma, P. Yang, Z. Liu, Y. Liu, H. Zhang, X. Dong, Y. Wang, K. Jiang, J. Hu, S. Nagasaki, K. Kitagawa, S. Calder, J. Yan, J. Sun, B. Wang, R. Zhou, Y. Uwatoko, J. Cheng, \textit{Bulk high-temperature superconductivity in pressurized tetragonal La$_2$PrNi$_2$O$_7$}, Nature \textbf{634}, 579-584 (2024).
\bibitem{dan_2024} Z. Dan, Y. Zhou, M. Huo, Y. Wang, L. Nie, M. Wang, T. Wu, and X. Chen, \textit{Spin-density-wave transition in double-layer nickelate La$_3$Ni$_2$O$_7$}, arXiv:2402.03952 (2024).
\bibitem{wang2024pressure} G. Wang, N. N. Wang, X. L. Shen, J. Hou, L. Ma, L. F. Shi, Z. A. Ren, Y. D. Gu, H. M. Ma, P. T. Yang, \textit{et al.}, \textit{Pressure-Induced Superconductivity in Polycrystalline La$_3$Ni$_2$O$_{7-\delta}$}, Phys. Rev. X \textbf{14}, 011040 (2024).
\bibitem{cai_2024} S. Cai, Y. Zhou, H. Sun, K. Zhang, J. Zhao, M. Huo, L. Nataf, Y. Wang, J. Li, J. Guo, \textit{et al.}, \textit{Low-temperature mean valence of nickel ions in pressurized La$_3$Ni$_2$O$_7$}, arXiv:2412.18343 (2024).
\bibitem{liu_2024} Z. Liu, M. Huo, J. Li, Q. Li, Y. Liu, Y. Dai, X. Zhou, J. Hao, Y. Lu, M. Wang, \textit{et al.}, \textit{Electronic correlations and partial gap in the bilayer nickelate La$_3$Ni$_2$O$_7$}, Nat. Commun. \textbf{15}, 7570 (2024).
\bibitem{YaoPRL2023} Zhihui Luo, Xunwu Hu, Meng Wang, Wei Wu, and Dao-Xin Yao, \textit{Bilayer Two-Orbital Model of La$_3$Ni$_2$O$_7$ under Pressure}, Physical Review Letters \textbf{131}, 126001 (2023).
\bibitem{Yang_2023_sPlusMinus} Q.-G. Yang, D. Wang, and Q.-H. Wang, \textit{Possible $s^\pm$-wave superconductivity in La$_3$Ni$_2$O$_7$}, Phys. Rev. B \textbf{108}, L140505 (2023).
\bibitem{Liu_2023_sPlusMinus} Y.-B. Liu, J.-W. Mei, F. Ye, W.-Q. Chen, and F. Yang, \textit{$s^\pm$-Wave Pairing and the Destructive Role of Apical-Oxygen Deficiencies in La$_3$Ni$_2$O$_7$ under Pressure}, Phys. Rev. Lett. \textbf{131}, 236002 (2023).
\bibitem{Yang_2023_ValenceBonds} Y.-F. Yang, G.-M. Zhang, and F.-C. Zhang, \textit{Interlayer valence bonds and two-component theory for high-$T_c$ superconductivity of La$_3$Ni$_2$O$_7$ under pressure}, Phys. Rev. B \textbf{108}, L201108 (2023).
\bibitem{Qin_2023_HighTc} Q. Qin and Y.-F. Yang, \textit{High-$T_c$ superconductivity by mobilizing local spin singlets and possible route to higher $T_c$ in pressurized La$_3$Ni$_2$O$_7$}, Phys. Rev. B \textbf{108}, L140504 (2023).
\bibitem{zhang_2023} Y. Zhang, L.-F. Lin, A. Moreo, and E. Dagotto, \textit{Electronic structure, dimer physics, orbital-selective behavior, and magnetic tendencies in the bilayer nickelate superconductor La$_3$Ni$_2$O$_7$ under pressure}, Phys. Rev. B \textbf{108}, L180510 (2023).
\bibitem{Christiansson_2023}Viktor Christiansson, Francesco Petocchi, and Philipp Werner, \textit{Correlated Electronic Structure of La$_3$Ni$_2$O$_7$ under Pressure}, Phys. Rev. Lett. \textbf{131}, 206501 (2023).
\bibitem{pupha_l2023} P. Puphal, P. Reiss, N. Enderlein, Y.-M. Wu, G. Khaliullin, V. Sundaramurthy, T. Priessnitz, M. Knauft, L. Richter, M. Isobe, \textit{et al.}, \textit{Unconventional crystal structure of the high-pressure superconductor La$_3$Ni$_2$O$_7$}, arXiv:2312.07341 (2023).
\bibitem{liao_2024} Z. Liao, Y. Wang, L. Chen, G. Duan, R. Yu, and Q. Si, \textit{Orbital-selective electron correlations in high-$T_c$ bilayer nickelates: from a global phase diagram to implications for spectroscopy}, arXiv:2412.21019 (2024).
\bibitem{zhang2024electronic} Y. Zhang, L.-F. Lin, A. Moreo, T. A. Maier, and E. Dagotto, \textit{Electronic structure, self-doping, and superconducting instability in the alternating single-layer trilayer stacking nickelates La$_3$Ni$_2$O$_7$}, Phys. Rev. B \textbf{110}, L060510 (2024).
\bibitem{zhang2023trends} Y. Zhang, L.-F. Lin, A. Moreo, T. A. Maier, and E. Dagotto, \textit{Trends in electronic structures and $s\pm$-wave pairing for the rare-earth series in bilayer nickelate superconductor R$_3$Ni$_2$O$_7$}, Phys. Rev. B \textbf{108}, 165141 (2023).
\bibitem{Tian_2024_Correlation} Y.-H. Tian, Y. Chen, J.-M. Wang, R.-Q. He, and Z.-Y. Lu, \textit{Correlation effects and concomitant two-orbital $s^\pm$-wave superconductivity in La$_3$Ni$_2$O$_7$ under high pressure}, Phys. Rev. B \textbf{109}, 165154 (2024).
\bibitem{Oh_2024_StrongPairing} H. Yang, H. Oh, and Y.-H. Zhang, \textit{Strong pairing from a small Fermi surface beyond weak coupling: Application to La$_3$Ni$_2$O$_7$}, Phys. Rev. B \textbf{110}, 104517 (2024).
\bibitem{Fan_2024_BilayerCoupling} Z. Fan, J.-F. Zhang, B. Zhan, D. Lv, X.-Y. Jiang, B. Normand, and T. Xiang, \textit{Superconductivity in nickelate and cuprate superconductors with strong bilayer coupling}, Phys. Rev. B \textbf{110}, 024514 (2024).
\bibitem{RyeePRL2024CMDFT} Siheon Ryee, Niklas Witt, and Tim O. Wehling, \textit{Quenched Pair Breaking by Interlayer Correlations as a Key to Superconductivity in La$_3$Ni$_2$O$_7$}, Phys. Rev. Lett. \textbf{133}, 096002 (2024).
\bibitem{wang2024electronic} Y. Wang, K. Jiang, Z. Wang, F.-C. Zhang, and J. Hu, \textit{The electronic and magnetic structures of bilayer La$_3$Ni$_2$O$_7$ at ambient pressure}, arXiv:2401.15097 (2024).
\bibitem{chen2024electronic} X. Chen, J. Choi, Z. Jiang, J. Mei, K. Jiang, J. Li, S. Agrestini, M. Garcia-Fernandez, H. Sun, X. Huang, \textit{et al.}, \textit{Electronic and magnetic excitations in La$_3$Ni$_2$O$_7$}, Nat. Commun. \textbf{15}, 9597 (2024).
\bibitem{lu2024interlayer} C. Lu, Z. Pan, F. Yang, and C. Wu, \textit{Interlayer-coupling-driven high-temperature superconductivity in La$_3$Ni$_2$O$_7$ under pressure}, Phys. Rev. Lett. \textbf{132}, 146002 (2024).
\bibitem{sakakibara2024} H. Sakakibara, N. Kitamine, M. Ochi, and K. Kuroki, \textit{Possible high $T_c$ superconductivity in La$_3$Ni$_2$O$_7$ under high pressure through manifestation of a nearly half-filled bilayer Hubbard model}, Phys. Rev. Lett. \textbf{132}, 106002 (2024).
\bibitem{jiang2024high} K. Jiang, Z. Wang, and F.-C. Zhang, \textit{High-temperature superconductivity in La$_3$Ni$_2$O$_7$}, Chin. Phys. Lett. \textbf{41}, 017402 (2024).
\bibitem{zhang2024strong} J.-X. Zhang, H.-K. Zhang, Y.-Z. You, and Z.-Y. Weng, \textit{Strong pairing originated from an emergent $Z_2$ Berry phase in La$_3$Ni$_2$O$_7$}, Phys. Rev. Lett. \textbf{133}, 126501 (2024).
\bibitem{Zheng_2025_TwoOrbital} Y.-Y. Zheng and W. Wu, \textit{$s^\pm$-wave superconductivity in the bilayer two-orbital Hubbard model}, Phys. Rev. B \textbf{111}, 035108 (2025).
\bibitem{Ko_LNO327AmbientSC} E. Ko, Y. Yu, Y. Liu, L. Bhatt, J. Li, V. Thampy, C. Kuo, B. Wang, Y. Lee, K. Lee, J. Lee, B. H. Goodge, D. A. Muller, H. Y. Hwang, \textit{Signatures of ambient pressure superconductivity in thin film La$_3$Ni$_2$O$_7$}, Nature (2024).DOI:10.1038/s41586-024-08525-3.
\bibitem{Zhou_2024_AmbientSC} G. Zhou, W. Lv, H. Wang, Z. Nie, Y. Chen, Y. Li, H. Huang, W. Chen, Y. Sun, Q.-K. Xue, Z. Chen, \textit{Ambient-pressure superconductivity onset above 40 K in bilayer nickelate ultrathin films}, arXiv:2412.16622v1 (2024).
\bibitem{liu2025superconductivity} Y. Liu, E. K. Ko, Y. Tarn, L. Bhatt, B. H. Goodge, D. A. Muller, S. Raghu, Y. Yu, and H. Y. Hwang, \textit{Superconductivity and normal-state transport in compressively strained La$_2$PrNi$_2$O$_7$ thin films}, arXiv:2501.08022 (2025).
\bibitem{XJZhou_ARPES_327} J. Yang, H. Sun, X. Hu, Y. Xie, T. Miao, H. Luo, H. Chen, B. Liang, W. Zhu, G. Qu, C.-Q. Chen, M. Huo, Y. Huang, S. Zhang, F. Zhang, F. Yang, Z. Wang, Q. Peng, H. Mao, G. Liu, Z. Xu, T. Qian, D.-X. Yao, M. Wang, L. Zhao, X. J. Zhou, \textit{Orbital-dependent electron correlation in double-layer nickelate La$_3$Ni$_2$O$_7$}, Nat. Commun. \textbf{15}, 4373 (2024).
\bibitem{LiPeng_ARPES_327_Film} P. Li, G. Zhou, W. Lv, Y. Li, C. Yue, H. Huang, L. Xu, J. Shen, Y. Miao, W. Song, Z. Nie, Y. Chen, H. Wang, W. Chen, Y. Huang, Z.-H. Chen, T. Qian, J. He, Y.-J. Sun, Q.-K. Xue, Z. Chen, \textit{Photoemission evidence for multi-orbital hole-doping in superconducting La$_2.85$Pr$_0.15$Ni$_2$O$_7$/SrLaAlO$_4$ interfaces}, arXiv:2501.09255 (2025).
\bibitem{cRPA2004} F. Aryasetiawan, M. Imada, A. Georges, G. Kotliar, S. Biermann, and A. I. Lichtenstein, \textit{Frequency-dependent local interactions and low-energy effective models from electronic structure calculations}, Phys. Rev. B \textbf{70}, 195104 (2004).
\bibitem{Scalapino_1986} Scalapino, D. J., Loh, E. and Hirsch, J. E. \textit{$d$-wave pairing near a spin-density-wave instability}, \textit{Phys. Rev. B} \textbf{34}, 8190--8192 (1986).
\bibitem{kamihara2008iron} Y. Kamihara, T. Watanabe, M. Hirano, and H. Hosono, \textit{Iron-based layered superconductor La[O$_{1-x}$F$_x$]FeAs ($x=0.05$--$0.12$) with $T_c=26$ K}, J. Am. Chem. Soc. \textbf{130}, 3296--3297 (2008).
\bibitem{qazilbash2009electronic} M. M. Qazilbash, J. J. Hamlin, R. E. Baumbach, L. Zhang, D. J. Singh, M. B. Maple, and D. N. Basov, \textit{Electronic correlations in the iron pnictides}, Nat. Phys. \textbf{5}, 647--650 (2009).
\bibitem{zhou2010electron} S. Zhou and Z. Wang, \textit{Electron correlation and spin density wave order in iron pnictides}, Phys. Rev. Lett. \textbf{105}, 096401 (2010).
\bibitem{yi2013observation} M. Yi, D. H. Lu, R. Yu, S. C. Riggs, J.-H. Chu, B. Lv, Z. K. Liu, M. Lu, Y.-T. Cui, M. Hashimoto, \textit{et al.}, \textit{Observation of temperature-induced crossover to an orbital-selective Mott phase in A$_x$Fe$_{2-y}$Se$_2$ (A = K, Rb) superconductors}, Phys. Rev. Lett. \textbf{110}, 067003 (2013).
\bibitem{bickers1989} N. E. Bickers and D. J. Scalapino, \textit{Conserving approximations for strongly fluctuating electron systems. I. Formalism and calculational approach}, Ann. Phys. \textbf{193}, 206--251 (1989).
\bibitem{witt2021} N. Witt, E. G. C. P. Van Loon, T. Nomoto, R. Arita, and T. O. Wehling, \textit{Efficient fluctuation-exchange approach to low-temperature spin fluctuations and superconductivity: From the Hubbard model to Na$_x$CoO$_2${\textperiodcentered}$y$H$_2$O}, Phys. Rev. B \textbf{103}, 205148 (2021).
\bibitem{wannier90} A. A. Mostofi, J. R. Yates, G. Pizzi, Y.-S. Lee, I. Souz, D. Vanderbilt, and N. Marzari, \textit{An updated version of wannier90: A tool for obtaining maximally-localised Wannier functions}, Comput. Phys. Commun. \textbf{185}, 2309 (2014).
\bibitem{Pizzi_2020} G. Pizzi, V. Vitale, R. Arita, S. Blügel, F. Freimuth, G. Géranton, M. Gibertini, D. Gresch, C. Johnson, T. Koretsune, \textit{et al.}, \textit{Wannier90 as a community code: new features and applications}, J. Phys.: Condens. Matter \textbf{32}, 165902 (2020).
\bibitem{Yue_327_data} C. Yue {\it et al.} https://zenodo.org/records/14634336 
\bibitem{Georges1996} A. Georges, G. Kotliar, W. Krauth, and M. J. Rozenberg, \textit{Dynamical mean-field theory of strongly correlated fermion systems and the limit of infinite dimensions}, Rev. Mod. Phys. \textbf{68}, 13 (1996).
\bibitem{Kotliar2006} G. Kotliar, S. Y. Savrasov, K. Haule, V. S. Oudovenko, O. Parcollet, and C. A. Marianetti, \textit{Electronic structure calculations with dynamical mean-field theory}, Rev. Mod. Phys. \textbf{78}, 865 (2006).
\bibitem{Werner2006} P. Werner, A. Comanac, L. de' Medici, M. Troyer, and A. J. Millis, \textit{Continuous-Time Solver for Quantum Impurity Models}, Phys. Rev. Lett. \textbf{97}, 076405 (2006).
\bibitem{Gull2011} E. Gull, A. J. Millis, A. I. Lichtenstein, A. N. Rubtsov, M. Troyer, and P. Werner, \textit{Continuous-time Monte Carlo methods for quantum impurity models}, Rev. Mod. Phys. \textbf{83}, 349 (2011).
\bibitem{Jarrell_1996_Bayesian} M. Jarrell and J. E. Gubernatis, \textit{Bayesian inference and the analytic continuation of imaginary-time quantum Monte Carlo data}, Phys. Rep. \textbf{269}, 133--195 (1996).
%
%
\bibitem{vasp_ref1} Kresse, G. and Hafner, \textit{Ab initio molecular dynamics for liquid metals}, J., Phys. Rev. B \textbf{47,} 558 (1993).
\bibitem{vasp_ref2} G. Kresse and J. Furthmüller, \textit{Efficiency of ab-initio total energy calculations for metals and semiconductors using a plane-wave basis set}, Computational Materials Science \textbf{6}, 15 (1996).
\bibitem{vasp_ref3} G. Kresse and J. Furthmüller, \textit{Efficient iterative schemes for ab initio total-energy calculations using a plane-wave basis set}, Phys. Rev. B \textbf{54}, 11169 (1996).
\bibitem{PBE_ref} J. P. Perdew, K. Burke, and M. Ernzerhof, \textit{Generalized gradient approximation made simple}, Phys. Rev. Lett. \textbf{77}, 3865 (1996).
\bibitem{Yue_327_data} C. Yue {\it et al.} https://zenodo.org/records/14634336 

\end{thebibliography}
\end{document}